\documentclass[aps,prstab,twocolumn,groupedaddress,longbibliography,nofootinbib]{revtex4-1}
\usepackage{paralist}
\usepackage{graphicx}
\usepackage{epstopdf}
\usepackage{epsfig}
\usepackage{natbib}
\usepackage{textcase}
\usepackage{bm}
\usepackage{url}
\usepackage{makecell}
\usepackage{multirow}
\usepackage{textgreek}
\usepackage{amsmath}
\usepackage{siunitx}

\begin{document}

    \title{Beam impact tests of a prototype target for the Beam Dump Facility at CERN: experimental setup and preliminary analysis of the online results}

\author{E.~Lopez~Sola}
\email[]{edmundo.lopez.sola@cern.ch}
\affiliation{CERN, 1211 Geneva 23, Switzerland}

\author{M.~Calviani}
\author{O.~Aberle}
\author{C.~Ahdida}
\author{P.~Avigni}
\author{M.~Battistin}
\author{L.~Bianchi}
\author{S.~Burger}
\author{J.~Busom~Descarrega}
\author{J.~Canhoto~Espadanal}
\author{E.~Cano-Pleite}
\author{M.~Casolino}
\author{M.A.~Fraser}
\author{S.~Gilardoni}
\author{S.~Girod}
\author{J-L.~Grenard}
\author{D.~Grenier}
\author{M.~Guinchard}
\author{C.~Hessler}
\author{R.~Jacobsson}
\author{M.~Lamont}
\author{A.~Ortega~Rolo}
\author{M.~Pandey}
\author{A.~Perillo-Marcone}
\author{B.~Riffaud}
\author{V.~Vlachoudis}
\author{L.~Zuccalli}

\affiliation{CERN, 1211 Geneva 23, Switzerland}

\date{\today}

\begin{abstract}
The Beam Dump Facility (BDF) is a project for a new facility at CERN dedicated to high intensity beam dump and fixed target experiments. Currently in its design phase, the first aim of the facility is to search for Light Dark Matter and Hidden Sector models with the Search for Hidden Particles (SHiP) experiment. At the core of the facility sits a dense target/dump, whose function is to absorb safely the 400 GeV/c Super Proton Synchrotron (SPS) beam and to maximize the production of charm and beauty mesons. An average power of 300 kW will be deposited on the target, which will be subjected to unprecedented conditions in terms of temperature, structural loads and irradiation. In order to provide a representative validation of the target design, a prototype target has been designed, manufactured and tested under the SPS fixed-target proton beam during 2018, up to an average beam power of 50~kW, corresponding to 350~kJ per pulse. The present contribution details the target prototype design and experimental setup, as well as a first evaluation of the measurements performed during beam irradiation. The analysis of the collected data suggests that a representative reproduction of the operational conditions of the Beam Dump Facility target was achieved during the prototype tests, which will be complemented by a Post Irradiation Examination campaign during 2020.
\end{abstract}


\maketitle

\section{Introduction and motivations}

The Beam Dump Facility (BDF), presently in its design phase, is a project for a multi-purpose facility at the North Area of CERN. The new facility will be dedicated to fixed target and beam dump experiments profiting from the 400~GeV/c Super Proton Synchrotron (SPS) proton beam. The first user of the facility will be the Search for Hidden Particles (SHiP) experiment, aiming at exploring Hidden Sector models and searching for Light Dark Matter~\cite{EOI_SHiP,Alekhin_SHiP,Anelli_SHiP,Ahdida_2019}. A dense target/dump (described in detail in Ref.~\cite{LopezSola:2019sfp}) will be located at the core of the facility, with a double function: (i) absorbing safely and reliably the SPS high-intensity beam (acting as a dump); (ii) maximizing the production of charm and beauty hadron decays and photons, all of them being potential sources of very weakly coupled particles.

\begin{figure}
\centering
\includegraphics[width=0.45\textwidth]{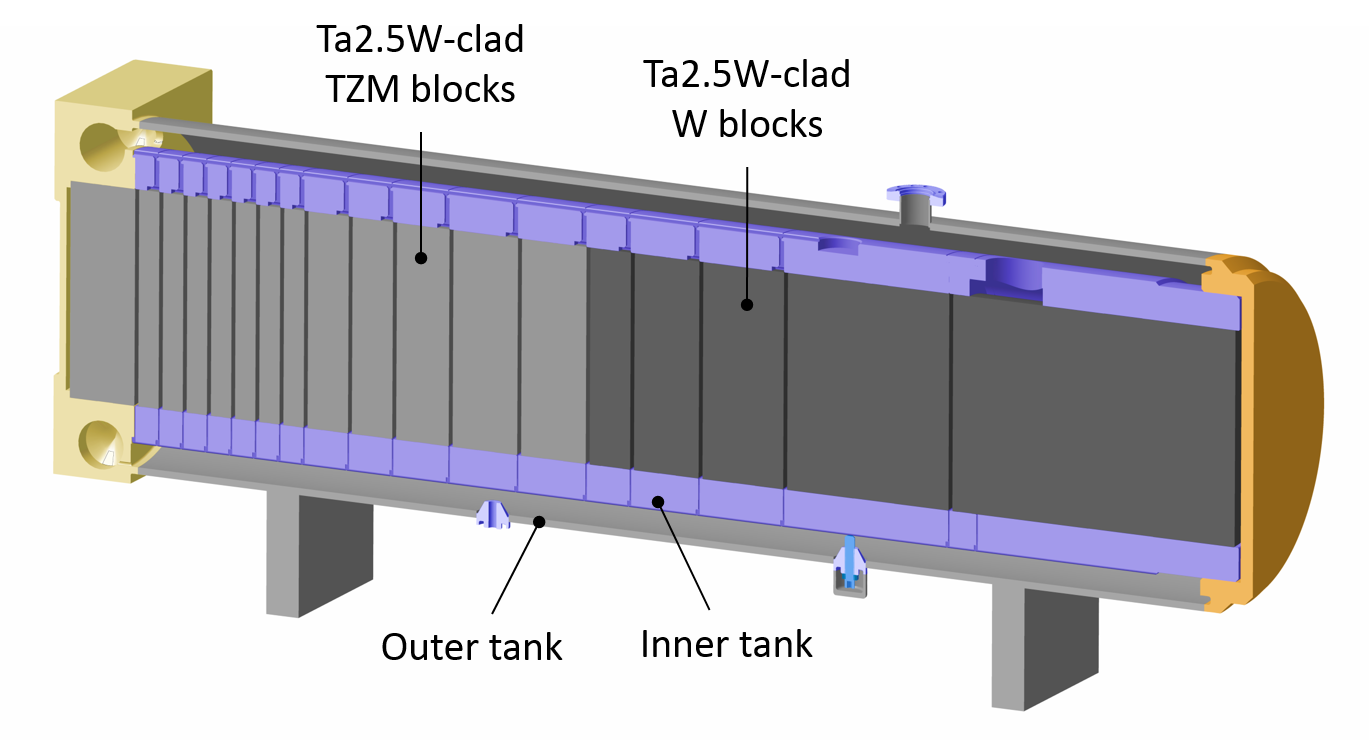}
 \caption{Current design of the BDF target assembly, as described in Ref.~\cite{LopezSola:2019sfp}. The target core materials (TZM and W) clad with a thin layer of Ta2.5W can be seen, as well as the two concentric tanks supporting the core and enclosing the target cooling system.}
\label{fig:finaltarget}
\end{figure}

The design of the BDF target and the corresponding target complex~\cite{BDFcomplex} is considered one of the most challenging aspects of the new facility, given the high levels of energy and power density that will be deposited during operation and the subsequent thermo-structural loads. The SPS proton beam is foreseen to impact the target during one second at an intensity of 4$\cdot10\textsuperscript{13}$ protons per pulse, followed by a cooling of 6.2 seconds. Out of the 355 kW average beam power impinging on target, about 300 kW will be deposited in the target assembly, while most of the remaining power will be dissipated in the surrounding steel and cast iron shielding. The target core design consists of several collinear cylinders of a molybdenum alloy (TZM) and pure tungsten, clad with a thin layer of a tungsten-containing tantalum alloy (Ta2.5W). The beam impacting on target will be diluted by the upstream magnets following a circular pattern, with a sweep frequency of 4 turns/s and a dilution radius of 50 mm~\cite{LopezSola:2019sfp}. Figure~\ref{fig:finaltarget} presents the current design of the BDF target. One of the most critical aspects of the target design is the cladding itself, that is expected to reach temperatures close to \SI{200}{\celsius} and cyclic stresses around 100~MPa. The BDF target is designed to withstand 5 years of operation for a total of $2\cdot10\textsuperscript{20}$ protons on target~\cite{LopezSola:2019sfp,BDFcomplex}.

Given the unprecedented regime of temperatures and stresses that are expected in the BDF target, a prototype of the target has been manufactured, assembled and tested during 2018. Representative beam characteristics were successfully reproduced, aiming at reaching similar operational scenarios in terms of temperature and stresses in the core, and at gaining experience on a reduced scale water cooling system.

The Beam Dump Facility conditions require slow extraction of the beam, therefore the HiRadMat facility at CERN~\cite{HRMT} could not be used, as done for target technologies tests in the past~\cite{PhysRevAccelBeams.21.073001,PhysRevAccelBeams.22.013401,doi:10.1002/mdp2.33,Davenne:2018nxs,PhysRevAccelBeams.21.033401,PhysRevSTAB.17.021004}. The target prototype experimental setup was therefore located in the North Area target zone of CERN (TCC2), where the SPS proton beam is regularly sent under slow extraction with intensities up to 3--4$\cdot10^{13}$ protons per pulse for physics experiments and test beams. Due to the configuration of the North Area, beam dilution is not available.

The main objectives of the target prototype tests are summarized hereafter:

\begin{itemize}
    \item Reproduce experimentally the level of temperatures and the magnitude of the thermal-induced stresses expected in the final target despite the lack of beam dilution;
    \item Evaluate the behaviour under thermal and structural cyclic loads of refractory clad materials, which will be subjected to temperature gradients of the order of \SI{100}{\celsius} per pulse during the final target operation;
    \item Cross-check the Finite Element Method (FEM) simulations performed. With that objective in mind, several target blocks were instrumented to perform online measurements;
    \item Explore the instrumentation survivability in challenging environments, including high levels of accumulated dose, high water speed and high pressure;
    \item Validate the performance of the target assembly cooling system and assess the effects of high cooling water velocity in contact with the blocks; 
    \item Perform detailed Post Irradiation Examination (PIE) studies after irradiation of the target prototype.
\end{itemize}

\section{Experimental setup}

\subsection{Overall layout of the area}

\begin{figure*}
\centering
\includegraphics[width=0.9\textwidth]{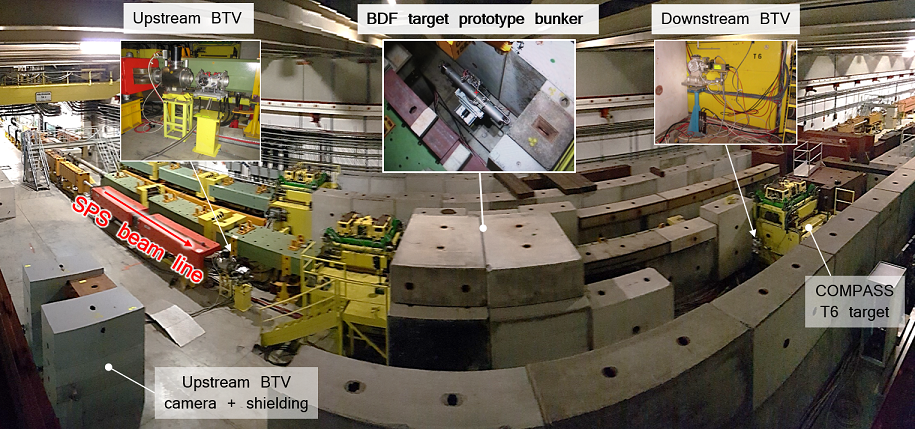}
 \caption{Top view of the TCC2 area showing the layout of the target prototype experimental setup: target prototype bunker, target beam screens (BTV), upstream BTV camera and camera shielding.}
\label{fig:T6layout}
\end{figure*}

The experimental setup of the BDF target prototype has been installed in the North Area target zone, upstream of the T6 beryllium production target, currently in use for the COMPASS experiment~\cite{COMPASS} (see Figure~\ref{fig:T6layout}). A new concrete shielding bunker was installed to house the experiment, as well as nearby cast iron shielding for the beam instrumentation.

Two dedicated beam screens (BTVs)~\cite{Bravin:2005vb} were installed, one upstream and one downstream of the experimental setup, in order to measure the beam profile impacting on the target and to align the beam with the target axis. The upstream BTV and its camera are shown in Figure~\ref{fig:T6layout}, the digital camera is surrounded by iron shielding to protect it from single event upsets and from total ionizing dose. 

A dedicated closed-circuit water cooling system was installed to avoid any possible contamination to the North Area primary cooling circuit in the event of target cladding failure. A full flow mixed-bed ion exchanger was also added to the loop.

\subsection{Prototype target core}

The target prototype tested in the North Area is a reduced scale replica of the final BDF target. The target prototype blocks have the same length distribution as the final BDF target, but a reduced diameter of 80 mm (instead of 250 mm). The materials used for the target prototype core are equivalent to the ones of the final target, with the core of the first 13 blocks made out of TZM (0.08\% titanium - 0.05\% zirconium - molybdenum alloy), and the last 5 blocks made out of pure tungsten (W).

In the BDF final target, it is foreseen to clad all the target blocks with a 1.5~mm-thick layer of a tungsten-containing tantalum alloy, Ta2.5W (2.5\% tungsten - tantalum alloy), in order to avoid undesired corrosion-erosion effects on the TZM or tungsten. Ta2.5W presents higher strength at high temperatures than pure tantalum and it is required given the temperature and stresses reached in the target blocks~\cite{LopezSola:2019sfp}. The use of Ta2.5W as cladding material is novel for a production target, while pure tantalum has already proven to be a reliable cladding material for tungsten blocks in other facilities (LANSCE, KENS and ISIS neutron source)~\cite{HIP1,HIP2,ISISclad}. Therefore, in the BDF target prototype both pure tantalum and Ta2.5W were used as cladding materials to compare their performance under beam irradiation~\cite{HIP_Busom}.

\begin{figure}[ht]
\resizebox{0.45\textwidth}{!}{
\includegraphics{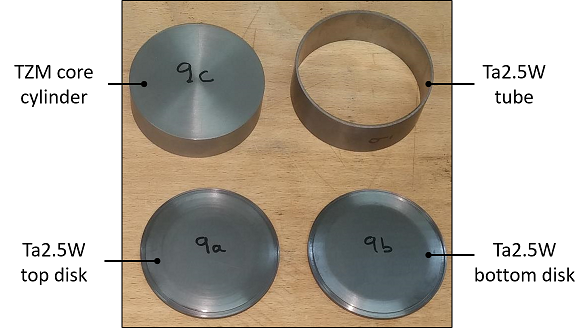}}
\caption{\label{fig:TGT:Block_exploded} View of the refractory metal parts required for the production of a Ta2.5W-clad TZM target block (80 ~mm diameter, 25~mm length) for the BDF target prototype.}
\end{figure}

The target blocks are made out of two different parts (see Figure~\ref{fig:TGT:Block_exploded}): 1) A TZM or W cylinder with a diameter of 77~mm and of different length according to the block position in the target core, and 2) a cladding made out of Ta or Ta2.5W, which encloses the TZM or W cylinder, and consists of a 1.5~mm-thick tube and two disks of 1.5~mm thickness. The materials were produced following the same manufacturing route as foreseen for the final target: the TZM material was obtained by means of multi-axial forging, while the W cylinders were produced via sintering and HIPing; all the Ta and Ta2.5W tubes were obtained by rolling, and the Ta and Ta2.5W disks were forged.

The core and cladding materials were joined via diffusion bonding achieved by means of Hot Isostatic Pressing (HIP), using a manufacturing process identical to the one of the final target blocks~\cite{LopezSola:2019sfp,HIP_Busom}. A summary of the materials and dimensions of the target prototype blocks is given in Table~\ref{tab:TGT:proto_dimensions}.

\begin{table}[htbp]
\centering
\caption{\label{tab:TGT:proto_dimensions} BDF target prototype blocks description, including the core and cladding materials used as well as the total length and weight of each block.}
\smallskip
\begin{ruledtabular}
\begin{tabular}{ccccc}
Block & Core & Cladding & Length & Weight \\
number & material & material & (mm) &(kg) \\
\hline
1                     & TZM                    & Ta                         & 80 & 4.1                  \\
2                     & TZM                    & Ta2.5W                     & 25 &1.3                  \\
3                     & TZM                    & Ta2.5W                     & 25 &1.3                  \\
4                     & TZM                    & Ta2.5W                     & 25&1.3                   \\
5                     & TZM                    & Ta2.5W                     & 25 &1.3                  \\
6                     & TZM                    & Ta2.5W                     & 25&1.3                   \\
7                     & TZM                    & Ta2.5W                     & 25&1.3                   \\
8                     & TZM                    & Ta                         & 25 &1.3                  \\
9                     & TZM                    & Ta2.5W                     & 50 &2.6                  \\
10                    & TZM                    & Ta                         & 50 &2.6                     \\
11                    & TZM                    & Ta                         & 65 &3.3                  \\
12                    & TZM                    & Ta                         & 80 &4.1                  \\
13                    & TZM                    & Ta                         & 80      &4.1             \\
14                    & W                      & Ta                         & 50 &4.7                  \\
15                    & W                      & Ta                         & 80      &7.5             \\
16                    & W                      & Ta                         & 100         &9.4         \\
17                    & W                      & Ta                         & 200             &18.8     \\
18                    & W                      & Ta                         & 350       &32.9          \\
\end{tabular}
\end{ruledtabular}
\end{table}

The pure tungsten blocks are all clad with pure tantalum and not with Ta2.5W. At the design stage of the target prototype, a good mechanical and chemical bonding between tungsten and Ta2.5W had not been produced via the HIP process. A successful bonding was later on achieved by adapting the HIP parameters employed~\cite{HIP_Busom}, validating the use of Ta2.5W as cladding material for all the target blocks in the final BDF target. 

\subsection{Prototype target assembly mechanical design}

\begin{figure}
\centering %
\includegraphics[width=0.45\textwidth]{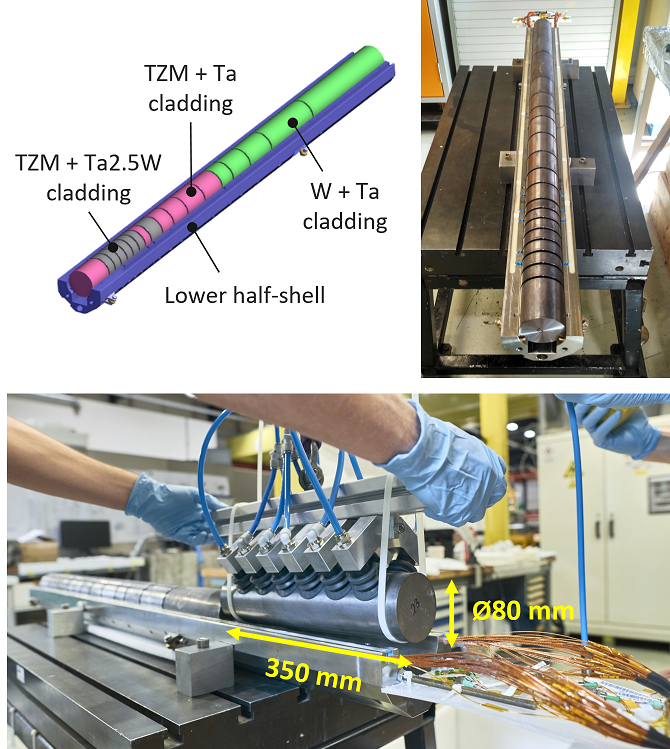}
\caption{\label{fig:TGT:proto_inner_tank} BDF target prototype inner tank lower shell and target blocks description (top left). Full target core assembly laying on the inner tank lower shell (top right). Installation of the largest tungsten core block ($\sim33$ kg) on the inner tank lower shell with help of a custom built suction cup (bottom)~\cite{CDS_pictures1}.}
\end{figure}

The target prototype assembly includes two concentric stainless steel tanks, similarly to the BDF final target. The inner tank consists of two half-shells, their function being to support the target blocks and to enclose the prototype cooling circuit. The target blocks are constrained in the radial direction by the two inner tank half-shells; in the beam axis direction, several pins are placed in order to allow free-body expansion of the blocks within 50~\textmu m, while ensuring a gap of 5~mm between the blocks, necessary for the water cooling circuit (see Section~\ref{Sec:FEM:CFD}). Figure~\ref{fig:TGT:proto_inner_tank} details the target blocks material distribution, and illustrates the target blocks during and after installation in the inner tank lower shell. 

\begin{figure}
\centering
\includegraphics[width=0.45\textwidth]{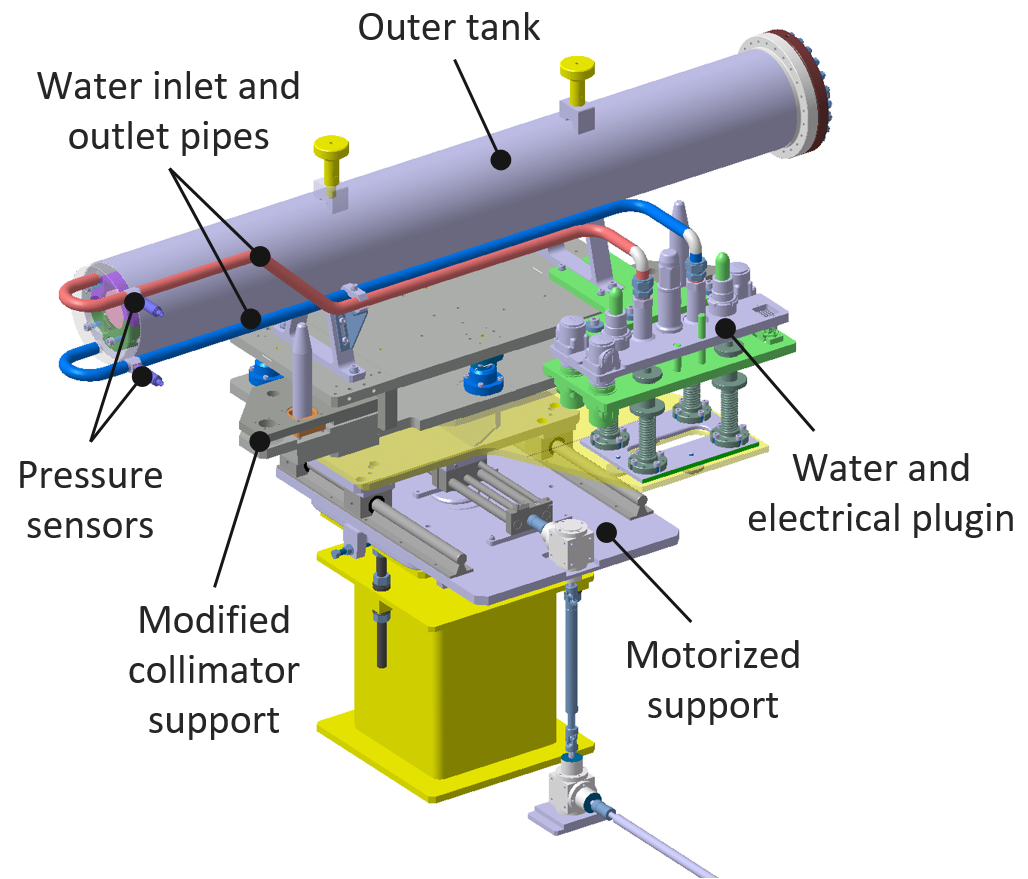}
\caption{Overview of the BDF target prototype assembly: the outer tank, the modified collimator support, the motorized support and the remote plug-in system are shown.}
\label{fig:fig:TGT:Proto:install}
\end{figure}

The outer stainless steel tank encloses the inner tank, ensuring the leak-tightness of the assembly, and also provides an interface for the electrical and water connections. The target prototype outer tank is placed on a modified Large Hadron Collider (LHC) collimator support~\cite{Collimator}, which consists of two different plates (upper and lower) with guiding pins. The design of the collimator support with plug-in features permits the precise installation and alignment of the upper plate on top of the lower plate by means of fully remote handling equipment without human intervention.

In addition, a dedicated additional plug-in system was developed and built in order to allow the remote connection and disconnection of the water and electrical connectors. The specific requirements of high water flow, high pressure and high radiation levels led to a fully metallic plug-in system compatible with the tele-manipulation tools of the CERN robotic team. The target prototype assembly is supported by a motorized table that allows its movement in the horizontal plane: the target prototype can be aligned with the beam axis for the beam tests, and removed from the beam after the tests in order to guarantee continuing physics operation for the COMPASS experiment. 

Due to the limited access to the experimental area during the installation period and the high dose rate expected after irradiation of the target prototype, the target prototype installation, replacement and removal was performed by only using the remotely manipulated crane of the TCC2 target zone and the telemanipulated CERN robots. Figure~\ref{fig:fig:TGT:Proto:install} describes the different components of the target prototype assembly and illustrates the target prototype remote handling with the TCC2 overhead travelling crane. 

\section{Thermo-mechanical and CFD calculations}
\label{Sec:FEM}

\subsection{Target prototype beam parameters}

The target prototype was tested in TCC2 using the same cycle configuration as for the final BDF target, i.e. spill length of 1.0 seconds and repetition rate of 7.2 seconds. The beam dilution foreseen for the final BDF target (four circular turns during the one second spill) could not be reproduced due to the lack of dilution magnets in the North Area transfer line. Therefore, the target prototype was tested under a non-diluted proton beam.

In consequence, the required beam intensity to reach representative temperatures and stresses with respect to the final target is lower and was estimated to be in the range of 3-4$\cdot10^{12}$ protons per pulse (ppp). The required beam spot size is also reduced with respect to the one expected during the final target operation, and is of the order of 3 x 2.5 mm. 
Table~\ref{tab:TGT:proto_beamparam} shows a comparison between the beam parameters of the BDF final target and the target prototype.

\begin{table}[hbtp]
\begin{ruledtabular}
\caption{\label{tab:TGT:proto_beamparam} The table summarizes the BDF final target and target prototype beam parameters~\cite{LopezSola:2019sfp}.}.
\begin{tabular}{lcc}
Baseline characteristics & \begin{tabular}[c]{@{}c@{}}BDF final\\ target\end{tabular}  & \begin{tabular}[c]{@{}c@{}}BDF target\\ prototype\end{tabular} \\
\hline
Proton momentum {[}GeV/c{]} & 400 & 400 \\
Beam intensity {[}p$^+$/cycle{]} & 4$\cdot10^{13}$ & 3--4$\cdot10^{12}$ \\
Beam dilution & Yes & No \\
Beam spot size (H/V) {[}mm{]} & 8/8 & 3/2.5 \\
Cycle length {[}s{]} & 7.2 & 7.2 \\
Spill duration {[}s{]} & 1.0 & 1.0 \\
Average beam power {[}kW{]} & 355 & 35 \\
Average power on target {[}kW{]} & 300 & 23 \\
\begin{tabular}[c]{@{}l@{}}Average beam power\\ during spill {[}MW{]}\end{tabular} & 2.56 & 0.26 \\
Power density per spill [MW/m$^3$] & 38 & 38\\
\end{tabular}
\end{ruledtabular}
\end{table}

\subsection{Thermal calculations}

The energy deposited on target by the SPS primary beam was calculated with the FLUKA Monte Carlo particle transport code~\cite{FLUKA_Code}, and imported into a Finite Element Analysis (FEA) software, ANSYS Mechanical\textsuperscript{\copyright}, for thermo-structural analysis. The heat transfer coefficient (HTC) distribution on the blocks surface was obtained my means of Computational Fluid Dynamics (CFD) calculations (see Section~\ref{Sec:FEM:CFD}), and was used as a boundary condition for the thermal analysis.

\begin{figure}
\centering %
\includegraphics[width=0.45\textwidth]{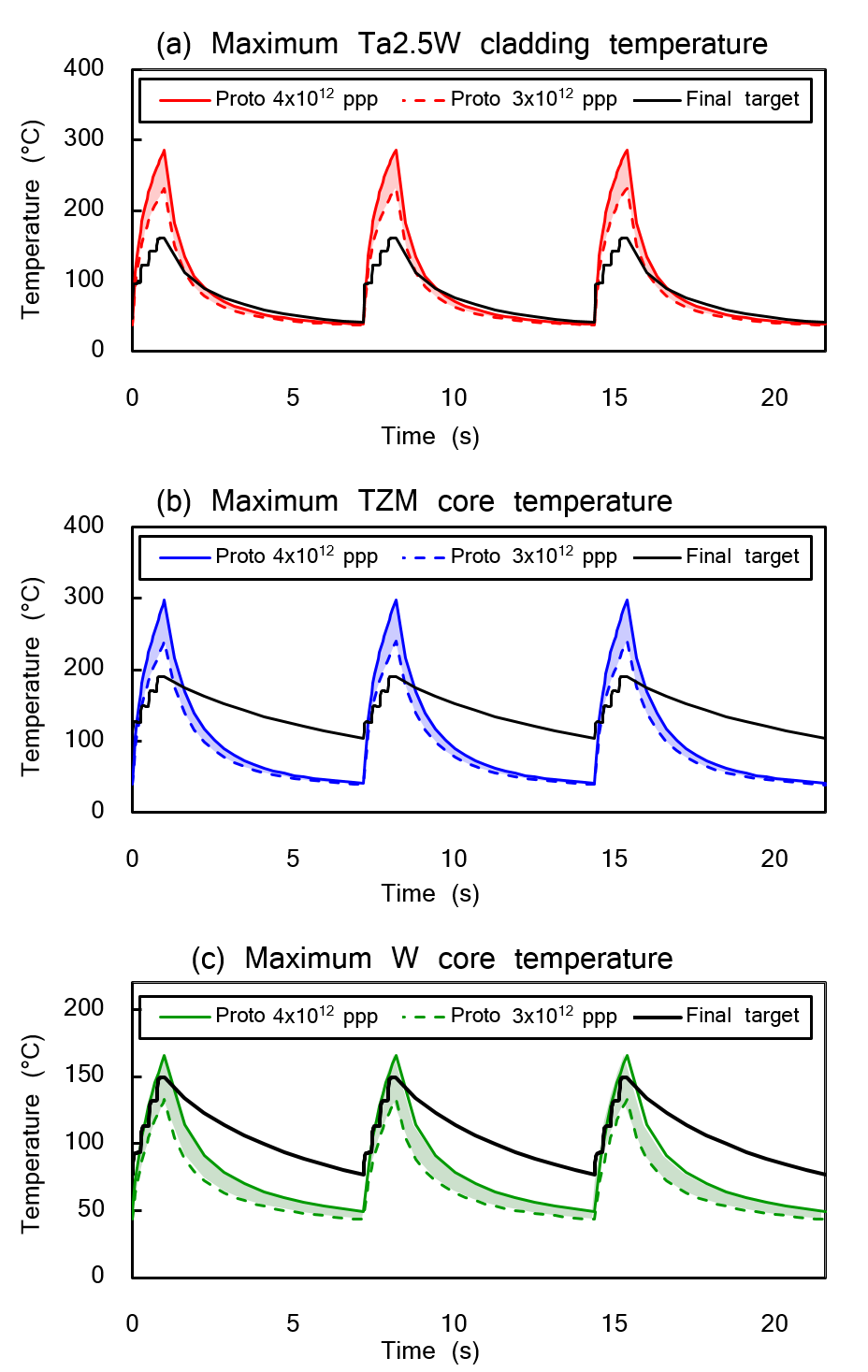}
\caption{\label{fig:TGT:proto_temps} Temperature evolution expected in the target materials during three beam pulses under steady-state regime, which is reached after around six pulses on target ($<$1 min operation). The temperature values are taken in the location of maximum temperature for the different target materials. Comparison between the final target and the target prototype operation under 3$\cdot10^{12}$ and 4$\cdot10^{12}$ protons per pulse (ppp).}
\end{figure}

A comparison between the maximum temperatures estimated in the target materials for the final target and the target prototype is shown in Figure~\ref{fig:TGT:proto_temps}. A band of intensities between 3$\cdot10^{12}$ ppp and 4$\cdot10^{12}$ ppp has been considered for the target prototype since the intensities reached during the beam tests were mainly comprised between these values. It can be seen that at the level of intensities of the prototype target beam tests, it is expected that the highest temperatures in the final BDF target materials are reached and even exceeded. 

\begin{figure}
\centering %
\includegraphics[width=0.45\textwidth]{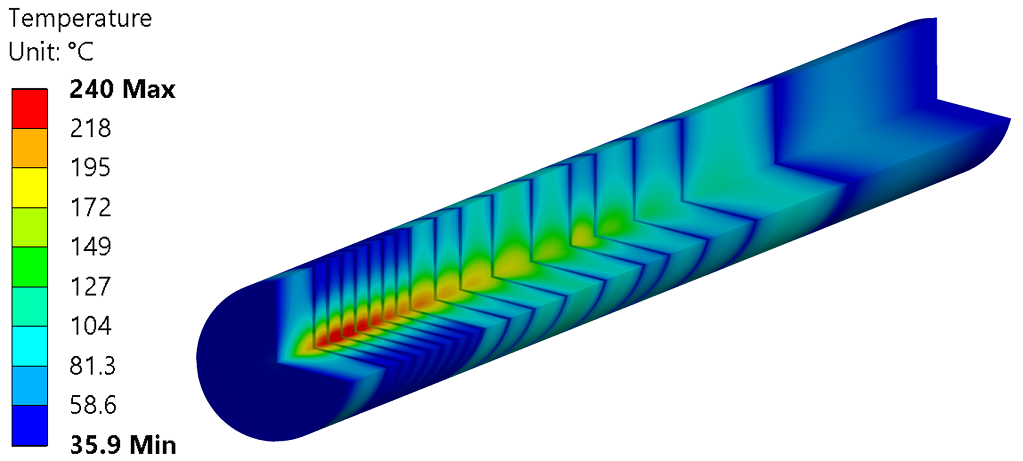}
\caption{\label{fig:TGT:proto_temp_all} Temperature distribution in the BDF target prototype core blocks after beam impact at 3$\cdot10^{12}$ ppp, after steady-state regime is reached. The maximum temperature expected is around \SI{240}{\celsius}, found in the TZM core of block 4.}
\end{figure}

Figure~\ref{fig:TGT:proto_temp_all} presents the FEA results of temperature distribution in the target prototype at the end of a beam impact at 3$\cdot10^{12}$ ppp. The temperature distribution differs from the one obtained in the BDF final target~\cite{LopezSola:2019sfp}, an effect which is due to the absence of beam dilution for the target prototype. Despite the different beam impact area in the target prototype and the final target, it is estimated that the stresses induced by the thermal loads have similar effects on the core/cladding interface.

\subsection{Structural calculations}

Structural simulations were carried out using as an input the calculated temporal evolution of temperature distribution. The thermal-induced stresses were considered as quasi-static, similarly to the final BDF target case, since the slow application of thermal loads (due to the pulse duration of one second) allows inertia effects to be neglected. A comparison between the maximum stress found in the different target materials for the BDF final target and for the prototype target test in a range of two different intensities is shown in Table~\ref{tab:TGT:proto_stress}. 

\begin{table}[htbp]
\centering
\caption{\label{tab:TGT:proto_stress} Maximum von Mises equivalent stress and maximum equivalent stress amplitude expected in the target materials for the final target~\cite{LopezSola:2019sfp} and for the target prototype in a range of two different intensities.}
\smallskip
\begin{ruledtabular}
\begin{tabular}{ccccc}
\multirow{3}{*}{{Material}} & \multicolumn{2}{c}{\begin{tabular}[c]{@{}c@{}}Maximum expected \\ stress {[}MPa{]}\end{tabular}} & \multicolumn{2}{c}{\begin{tabular}[c]{@{}c@{}}Maximum expected \\ $\sigma_{a,eq}$ {[}MPa{]}\end{tabular}} \\\cline{2-3}\cline{4-5}  & Final & Target prototype & Final & Target prototype\\
 & target & 3--4$\cdot10^{12}$ ppp & target & 3--4$\cdot10^{12}$ ppp\\
 \hline
TZM & 130 & 145 -- 195 & 60 & 70 -- 95 \\
W & 95 & 85 -- 110 & 30 & 25 -- 35 \\
Ta2.5W & 95 & 85 -- 120 & 45 & 40 -- 60\\                       \end{tabular}
\end{ruledtabular}
\end{table}

For fatigue considerations, the maximum equivalent stress amplitude $\sigma_{a,eq}$ expected under the final BDF target and the target prototype conditions is also presented. The value of equivalent stress amplitude has been computed from the amplitude of the stresses in the principal directions using the von Mises equation, and is expected to give the same fatigue life under uni-axial loading as the multi-axial state of stress found in the final target and the target prototype~\cite{LopezSola:2019sfp}.

Table~\ref{tab:TGT:proto_stress} shows that, in terms of maximum stress and maximum stress amplitude, the target prototype is subjected to even more challenging conditions than the BDF final target for the range of intensities (between 3$\cdot10^{12}$ and 4$\cdot10^{12}$ ppp) reached during the beam tests. 

\begin{figure}
\centering %
\includegraphics[width=0.45\textwidth]{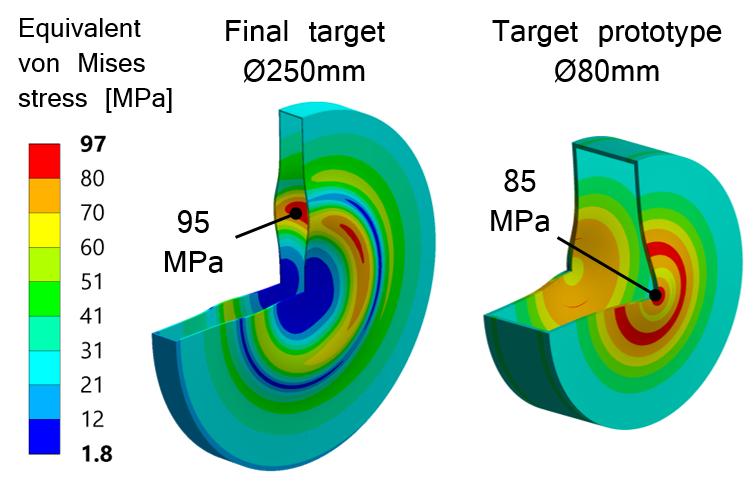}
\caption{\label{fig:TGT:proto_stress_distribution} Von Mises equivalent stress distribution in the Ta2.5W cladding  of the most loaded target block (block 4) after beam impact. Comparison between the final target (left) and the target prototype at 3$\cdot10^{12}$ ppp (right). The different stress distribution due to the beam dilution or non-dilution is noticeable.}
\end{figure}

\begin{figure}
\centering %
\includegraphics[width=0.45\textwidth]{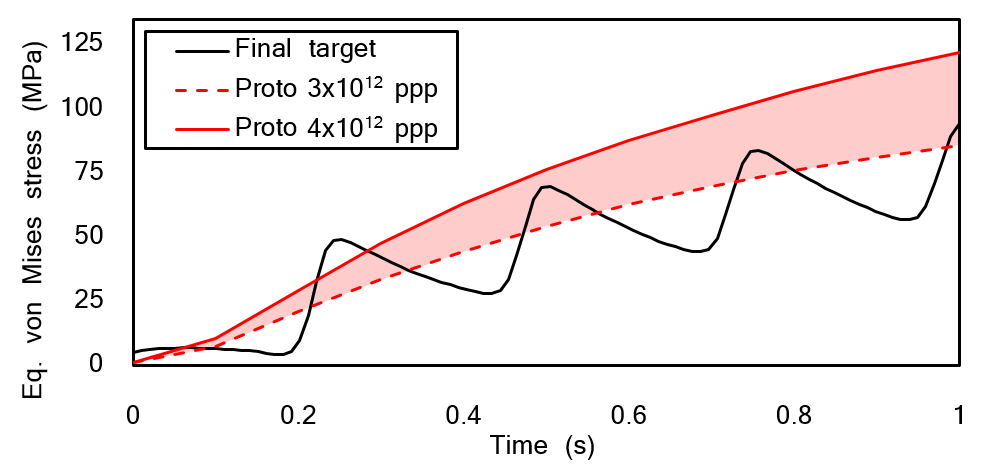}
\caption{\label{fig:TGT:proto_stress_evolution} Von Mises equivalent stress evolution at the point of maximum stress of the Ta2.5W cladding (block 4) during the beam impact of 1 second. Comparison between the final target and the target prototype at two different intensities. For the final target, the effect of the beam dilution in four circular turns can be clearly observed.}
\end{figure}

The fact that the beam is diluted for the final BDF target and not for the prototype leads to a different temperature distribution, and therefore, to a different stress field, as illustrated by Figure~\ref{fig:TGT:proto_stress_distribution}. The stress evolution is also dissimilar for both cases. Figure~\ref{fig:TGT:proto_stress_evolution} shows a comparison between the evolution of the von Mises equivalent stress in the Ta2.5W cladding of the final target and the target prototype. 

Despite the difficulties in reproducing an identical stress state in the prototype blocks, the maximum level of stress foreseen in the final target was reproduced and even exceeded in the target prototype tests, as will be shown in Section~\ref{Sec:TGT:Proto:Tests}.

\subsection{Prototype cooling system and Computational Fluid Dynamics (CFD) analysis}
\label{Sec:FEM:CFD}

The design of the target prototype cooling system aimed to provide an initial validation of the final target cooling system design as well as guaranteeing a representative heat transfer coefficient in the target prototype. The target prototype cooling design replicates the major characteristics of the BDF final target cooling system, including 1) water cooling at a high pressure of around 2.2$\cdot10^6$~Pa (22 bar), 2) 5~mm channels between the blocks for the water passage, 3) high water speed between plates (around 4~m/s) and 4) serpentine configuration of the water flow.

\begin{figure*}
\centering %
\includegraphics[width=1\textwidth]{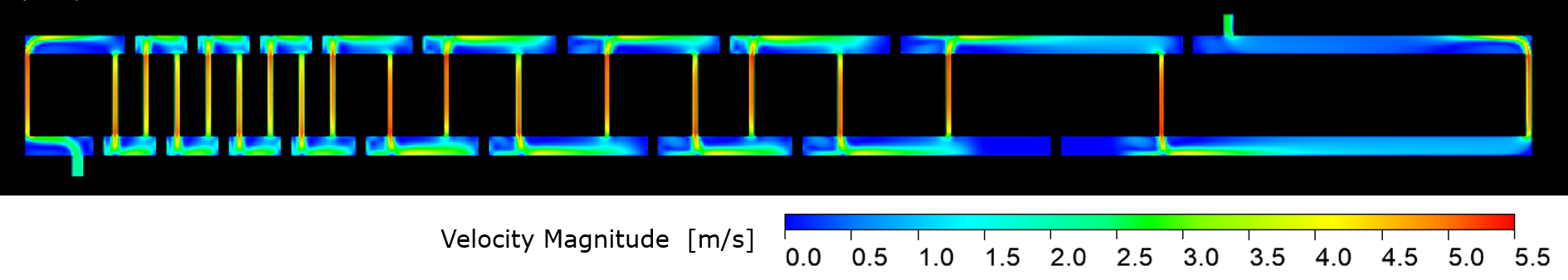}
\caption{\label{fig:TGT:proto_CFD_vel} 2D contour of velocity magnitude in the target prototype cooling system (serpentine configuration), obtained by means of CFD calculations. The water flow reaches high velocities of around 4 m/s in the 5~mm gap between the target blocks.}
\end{figure*} 

Several cooling circuit configurations were investigated in order to minimize the required mass flow rate while obtaining a uniform fluid velocity and high HTC in the channels. The single-channel (serpentine) configuration presented in Figure~\ref{fig:TGT:proto_CFD_vel} was found to be the optimal choice in terms of flow velocity uniformity and overall mass flow rate. This design differs from the final target cooling system design, which features two parallel streams following a serpentine circulation~\cite{LopezSola:2019sfp}.

The pressure drop along the target prototype cooling circuit estimated by means of CFD calculations with ANSYS CFX~\textsuperscript{\copyright} is around 2.5$\cdot10^5$~Pa (2.5 bar). Water cooling tests were performed prior to the target prototype installation in order to measure experimentally the pressure drop in the target prototype cooling circuit and to assess the leak tightness of the assembly. The target prototype was tested under static pressure of 3.2$\cdot10^6$~Pa (32 bar) and with circulating water at 2.2$\cdot10^6$~Pa (22 bar) and a flow rate of 1~kg/s, calculated analytically to obtain an average speed in the channels of 4~m/s. As an outcome of the cooling tests, the leak tightness of the assembly was validated, and the pressure drop was measured to be around 3$\cdot10^5$~Pa (3 bar), showing a fair agreement with the CFD calculations.

\begin{figure}
\centering %
\includegraphics[width=0.35\textwidth]{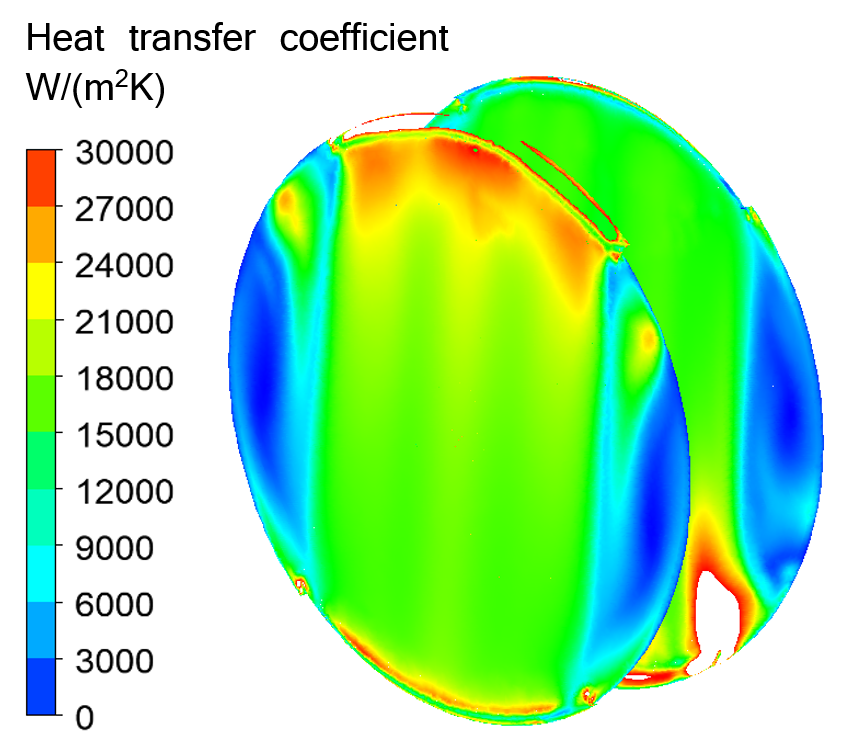}
\caption{\label{fig:TGT:proto_CFD_HTC} Heat Transfer Coefficient (HTC) distribution in the front and back surfaces of block~4. The average HTC found in the surfaces is around 15000~W/(m$^2$K).}
\end{figure}

The high water speed in the vertical channels is expected to result in a surface HTC of about 15000~W/(m$^2$K) in average. The HTC distribution obtained by CFD simulations is non-uniform, as shown in Figure~\ref{fig:TGT:proto_CFD_HTC} for block~4, and was imported as boundary condition for the FEM thermal simulations previously presented.

\section{Target prototype instrumentation}
\label{Sec:TGT:Proto:Instru}

The target blocks were instrumented to assess the thermal and structural response of the target materials under beam irradiation, and to compare the measured behaviour with the FEM simulations. Four blocks were instrumented, covering a combination of the different materials employed for the target prototype, as described in Table~\ref{tab:TGT:instrublocks}.

\begin{table}[hbtp]
\caption{Materials and dimensions of the four target prototype blocks that were selected to be instrumented.}
\label{tab:TGT:instrublocks}
\small
\begin{ruledtabular}
\begin{tabular}{ccccc}
\multirow{2}{*}{Block}& \multicolumn{2}{c}{Materials}& \multicolumn{2}{c}{Dimensions}   \\ \cline{2-3} \cline{4-5} 
number    & Core & Cladding & \makecell{Diameter\\ (mm)} & \makecell{Length\\(mm)} \\ \hline
4   & TZM & Ta2.5W & \multirow{4}{*}{80} & 25 \\
8  &  TZM & Ta &  & 25 \\
9  &  TZM & Ta2.5W &  & 50 \\ 
14  &  W & Ta &   & 50  \\ 
\end{tabular}
\end{ruledtabular}
\end{table}

The selected blocks are expected to be the most critical in terms of thermal and structural loads for the different materials: blocks 4 and 9 for Ta2.5W and TZM, block 8 for pure tantalum and block 14 for pure tungsten.

The purpose of the instrumentation was to measure the temperature, radial strain and circumferential strain in several points of the flat surfaces (upstream and downstream) of the blocks. The harsh working environment of the experiment (4 m/s water velocity, pressure of 2.2$\cdot10^6$~Pa (22 bar), interaction with the high-energy particle beam and subsequent accumulated dose rate) oriented the choice of sensors and services towards waterproof/watertight, radiation hard and pressure resistant equipment. The instrumentation is expected to be subjected to a total dose of 100 MGy integrated over the course of the experiment, estimated by means of FLUKA Monte Carlo simulations. Other restrictions were imposed by the capability to accurately measure the physical quantities of interest and by the specific design of the BDF target prototype (e.g. strain rate, available space for cabling, gap between blocks).

\begin{figure}
\centering %
\includegraphics[width=0.4\textwidth]{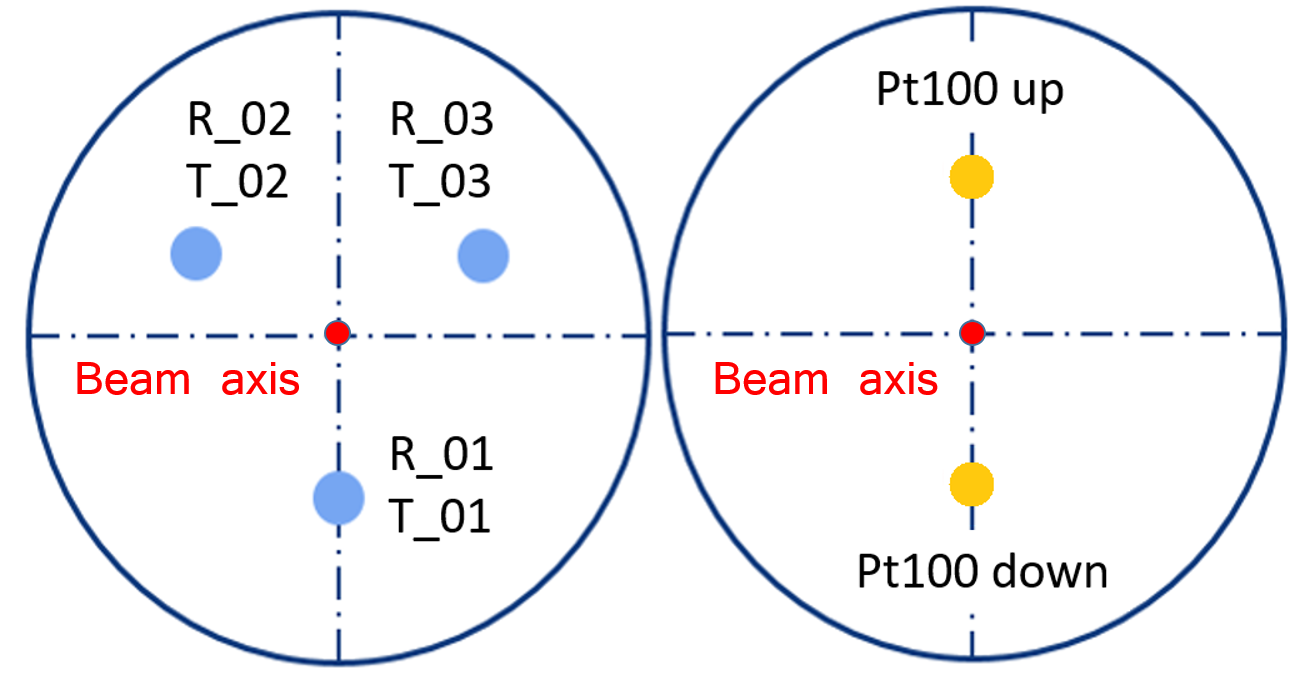}
\caption{\label{fig:TGT:sensors_position} Bi-axial strain gauges for radial (R) and transversal (T) measurements in three positions (01, 02 and 03) and temperature sensors (Pt100) in two positions. Design for the upstream (left) and downstream (right) faces of blocks 4, 8, 9 and 14. The expected beam impact position is also shown.}
\end{figure}

Taking into account the requirements and constraints of the target prototype experiment, the following measuring points were selected (see Figure~\ref{fig:TGT:sensors_position}): three measurement points at 120$^{\circ}$ in the upstream faces for radial and transversal strain (resistive bi-axial strain gauges), and two measurement points at 180$^{\circ}$ on the vertical axis of the downstream faces for temperature sensing (Pt100). The measuring points were placed at a distance of 20$\pm$0.5~mm from the target axis.

The selected strain gauges and temperature sensors - which are not waterproof themselves - were employed in combination with a protective cover agent to protect the electrical connections from the water stream~\cite{Ahdida:2650896}. In addition, a good bonding between the strain gauges and the block surface is necessary to minimize the strain dissipation, and an epoxy adhesive glue was used for that purpose. In order to minimize the error of the temperature measurements, an epoxy adhesive with high thermal conductivity was applied for the bonding of the Pt100 sensors.

A validation campaign was carried out prior to the beam tests to verify the suitability of the measuring equipment, as very little technical information was available at the BDF target prototype operational conditions. An experimental setup was designed and manufactured to test strain gauges, glues, protective agents and electrical feedthroughs under pressurized water and high flow rate. The test ran for several days under a circulating water flow of 4 m/s at 2.2$\cdot10^6$~Pa (22 bar). The test-rig operation proved that the strain gauges were successfully protected by the cover agents employed, that the underwater environment did not influence the measurements, and that the electrical feedthroughs could withstand the operational pressure without leakages. Therefore, the use of the selected instrumentation for the target prototype tests was validated. 

\begin{figure}
\centering %
\includegraphics[width=0.4\textwidth]{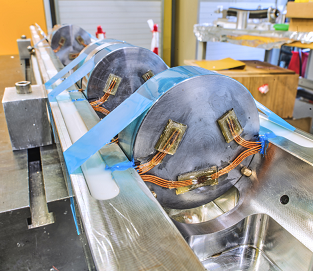}
\caption{\label{fig:TGT:proto_assy_instru} Instrumented prototype blocks installed on the inner tank lower half shell. The three bi-axial strain gauges at 120$^{\circ}$ covered by the protective agent can be observed~\cite{CDS_pictures1}.}
\end{figure}

The beam instrumentation worked successfully during the tests, and most of the temperature sensors and strain gauges survived during the three days of operation under beam despite the harsh environmental conditions. The instrumented target prototype blocks placed on the inner tank lower half-shell are shown in Figure~\ref{fig:TGT:proto_assy_instru}. 

\section{Beam tests}
\label{Sec:TGT:Proto:Tests}

The BDF target prototype tests under beam took place over three different days. The target prototype operated under the BDF cycle for several hours during the execution of the tests. Other repetition rates were also employed depending on the availability of the beams. In total, more than 14 hours of beam were dedicated to the target prototype tests. A total of around 2.4$\cdot10^{16}$ protons on target (POT) was reached, with the target being subjected to almost 10$^{4}$ beam pulse cycles. 

The intensity delivered to the target during the execution of the beam tests varied mostly between 3$\cdot10^{12}$ and 4$\cdot10^{12}$ protons per pulse, which corresponds to an average beam power between 27 and 35 kW under the BDF cycle. The maximum intensity reached during the tests was around 5.5$\cdot10^{12}$ ppp (50 kW average power), corresponding to a beam energy of 350~kJ.

Out of the power delivered on target, almost 65\% is absorbed by the target assembly and dissipated by the water cooling system. It was possible to estimate this quantity by measuring the temperature increase of the cooling water in the circuit\footnote{As an example, for an intensity of 3.75$\cdot10^{12}$ ppp, which corresponds to an average power on target of 33 kW, the water temperature increase measured was around \SI{5}{\celsius}. This temperature rise is equivalent to an approximate power of 21 kW dissipated by the cooling system, which means that around 63\% of the power delivered was absorbed by the target assembly.}. This value is indeed coherent with the results of the FLUKA Monte Carlo calculations.

In this section, some of the post-processed measurements of pressure, temperature and strain recorded during the first day of beam tests are presented, as well as a comparison with the expected results from the FEM and CFD calculations. 

\subsection{Beam characteristics}

\begin{figure}
\centering %
\includegraphics[width=0.45\textwidth]{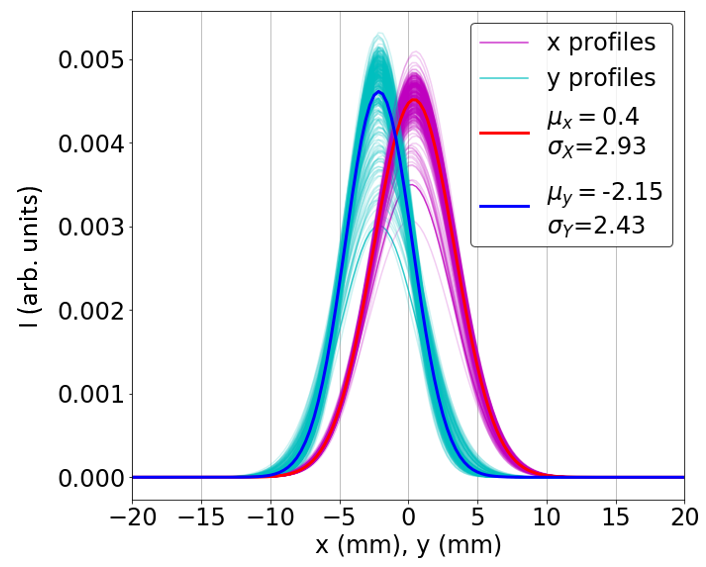}
\caption{\label{fig:TGT:btv_offset}Beam profile along the horizontal (x) and vertical (y) directions recorded by the upstream BTV during the beam tests at medium intensity (3.25$\cdot10^{12}$ ppp). A total of 750 beam pulses are presented, as well as the average beam profile in both directions. The beam centroid has an offset of about 2 mm down in the vertical direction. The beam spot size on target is around 2.9 x 2.4 mm 1$\sigma$, fairly comparable to the requested beam size (3 x 2.5 mm 1$\sigma$).}
\end{figure}

The beam steering and tuning to reach the target centre with the required beam spot dimensions was performed profiting from the beam instrumentation installed (one BTV upstream the target, another one downstream). Figure~\ref{fig:TGT:btv_offset} presents a superposition of the beam profile on target in the horizontal and vertical plane, extracted from the upstream BTV data recorded during the tests at medium intensity (3.25$\cdot10^{12}$ ppp), coupled to a gaussian fit. 

Good beam centering was achieved in the horizontal direction, while in the vertical direction the beam centroid had an offset of 2 to 3 mm towards the bottom with respect to the nominal position. This issue had an influence in the recorded data, as will be described in the following paragraphs. The jitter of the beam, calculated as the standard deviation of the different beam centroid positions recorded during a given period, was below 0.15 mm in both the horizontal and vertical directions throughout the tests. The beam size obtained during the tests was close to the requested size of 3 x 2.5 mm 1$\sigma$, as shown in Figure~\ref{fig:TGT:btv_offset}.

\subsection{Pressure and temperature measurements}
\label{Sec:TGT:Proto:Tests:temp}

The pressure measured by the sensors placed at the inlet and outlet pipes of the prototype target cooling circuit (see Figure~\ref{fig:fig:TGT:Proto:install}) showed a relatively constant pressure drop of 3$\cdot10^5$~Pa (3 bar), compatible with the values predicted by the CFD calculations presented in Section~\ref{Sec:FEM:CFD}. The maximum pressure varied between 2.1 and 2.2$\cdot10^6$~Pa (21-22 bar) during the beam tests, as a result of the pressure regulation at the supply and the beam-induced temperature changes.

\begin{figure}
\centering %
\includegraphics[width=0.44\textwidth]{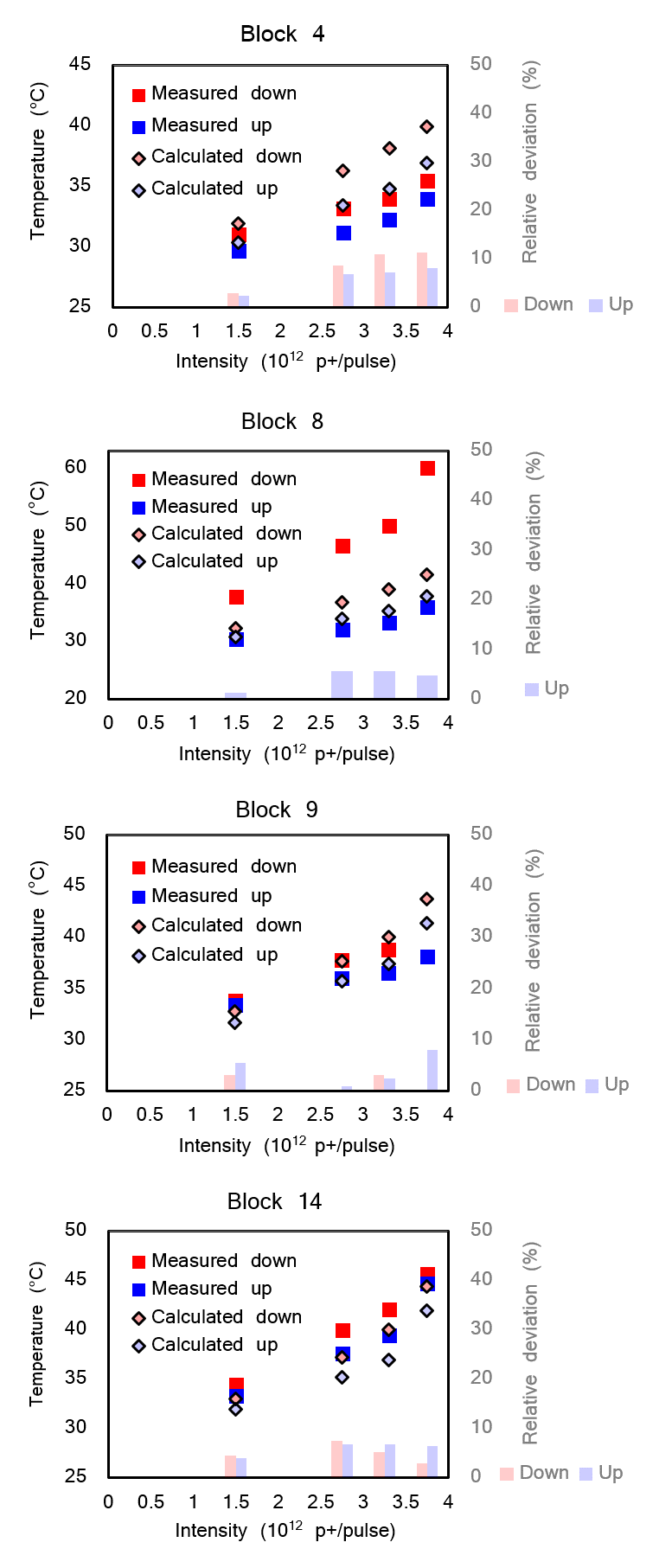}
\caption{\label{fig:TGT:proto_temp_vs_int} Maximum temperature measured by the two temperature sensors (up and down) of the four instrumented target blocks. The plotted values correspond to the average maximum temperature measured at different intensities once steady-state regime was reached: 1.5, 2.75, 3.25$\cdot10^{12}$ ppp (10.8 seconds cycle) and 3.75$\cdot10^{12}$ ppp (7.2 seconds cycle). A comparison with the temperatures obtained from the FEM simulations is shown, as well as the relative deviation between both values, calculated as $(T_{\text{meas}}-T_{\text{calc}})/T_{\text{calc}}.$}
\end{figure}

The beam intensity was increased throughout the execution of the tests, starting from 1.5$\cdot10^{12}$ ppp and reaching 3.75$\cdot10^{12}$ ppp at the end of the tests. The temperatures recorded by the Pt100 sensors increased for an increased beam intensity, as predicted by the FEM calculations. Figure~\ref{fig:TGT:proto_temp_vs_int} presents the maximum temperature measured in all the sensors for different beam intensities, as well as a comparison with the expected values from the thermal calculations. 

The maximum values of measured temperature are taken for a given intensity and a given repetition rate after steady-state regime was reached. The temperature measurements recorded during the tests presented oscillations of $\pm \SI{0.4}{\celsius}$, due to periodic fluctuations in the temperature of the water supply. The data presented in Figure~\ref{fig:TGT:proto_temp_vs_int} corresponds to the average maximum temperature taking into account the oscillations recorded. 

For the FEM thermal simulations, a beam offset of 2~mm towards the bottom of the target was taken into account. As shown by Figure~\ref{fig:TGT:proto_temp_vs_int}, the fact that the beam is displaced towards the bottom affects the temperature measurements: for all the blocks, the lower Pt100 sensors ("Pt100 down") measured higher values than the upper sensors ("Pt100 up").

The temperature recorded by the upper Pt100 of block number 8 presents a large deviation with respect to the simulated values, and is much higher than the temperatures measured in blocks 4 and 9, contrary to what was expected. This issue will need to be examined in the future after the target opening, and could be related to a reduced heat dissipation in the proximity of the sensor: the large amount of instrumentation (and the corresponding protective agents) placed in the 5~mm channel between blocks 8 and 9, which were both instrumented, could have a detrimental effect on the flow circulation, thus affecting the heat dissipation at some points of the block surface. The deviation between this temperature measurement and the FEM calculations results is above 20\%, but is not considered to be representative and has not been plotted. The upper sensor of block number 9 failed after several hours of operation, and no temperature was recorded at 3.75$\cdot10^{12}$~ppp.

Generally, the measured temperatures are coherent with the values predicted by the FEM calculations. The relative deviation between the calculated values and the measured ones is in most cases below 10\%. 

Several factors can justify the difference found between the measured and simulated results. These include, for example: 1) the systematic error associated to the energy deposition values from FLUKA Monte Carlo simulations; 2) the CFD calculations carried out to estimate the HTC do not take into account the influence of the Pt100 sensors, strain gauges, and their protective agents in the flow behaviour; 3) the thermal calculations do not take into account neither the thermal resistance created by the layer of glue used to bond the Pt100 to the block surface nor the reduction of heat dissipation due to the protective agent covering the sensors.

\begin{figure}
\centering %
\includegraphics[width=0.45\textwidth]{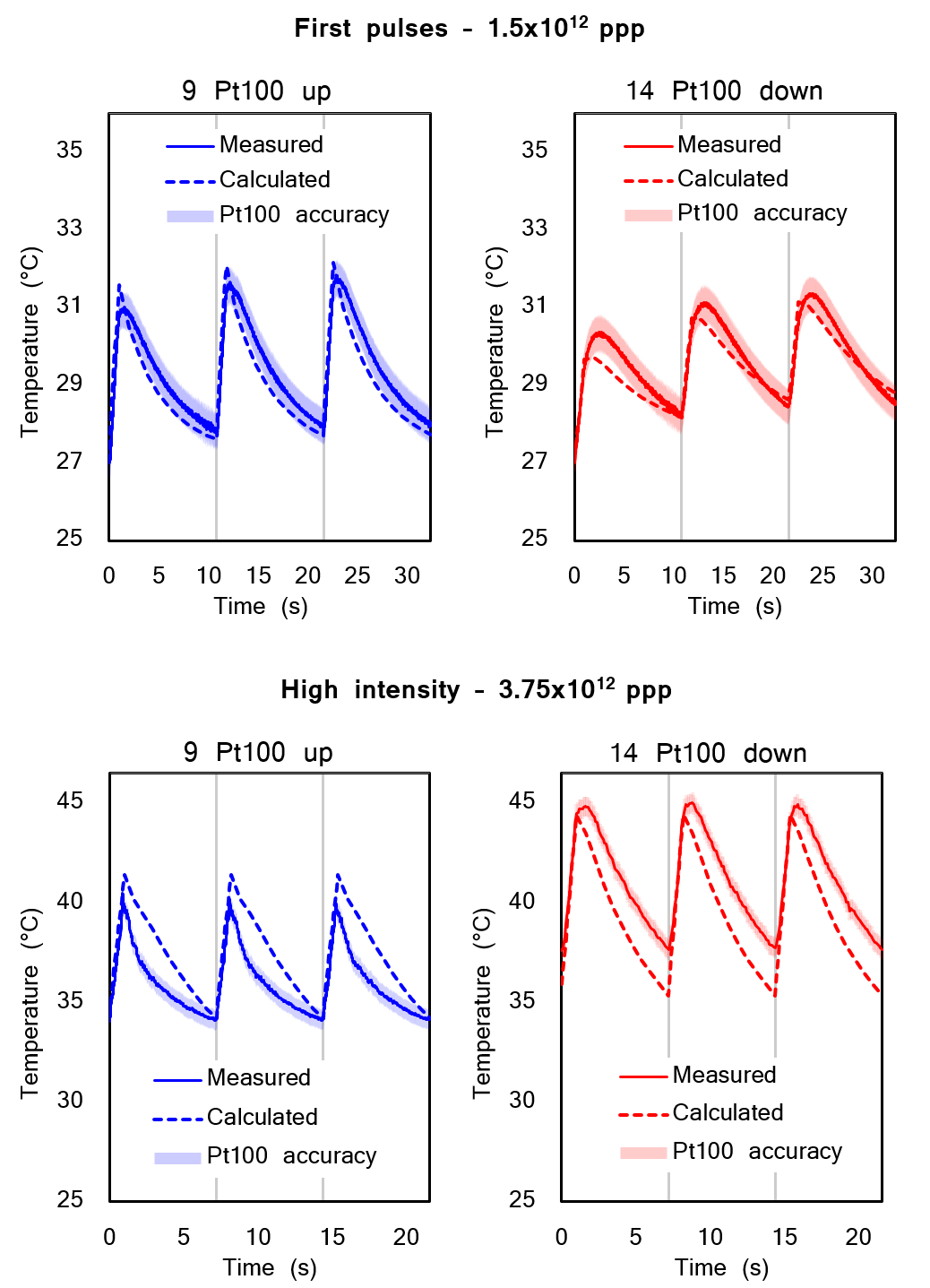}
\caption{\label{fig:proto_9_evo} Temperature measured by two sensors (upper Pt100 of block 9 and lower Pt100 of block 14) during three beam pulses at two different intensities: first beam pulses on target at an intensity of $1.5\cdot10^{12}$ ppp and a repetition rate of 10.8 seconds; $3.75\cdot10^{12}$ ppp after steady-state regime was reached under the BDF cycle. Pt100 accuracy calculated as $\pm(0,3 + 0,005*T)^{\circ}$C for a given temperature $T$ (class B accuracy). Comparison with the FEM simulations results under the same operational conditions.}
\end{figure}

For most of the measuring points, a good correlation is also found between the evolution of measured temperature with time and the evolution calculated via FEM simulations. Figure~\ref{fig:proto_9_evo} presents the temperature evolution during three beam pulses recorded by two Pt100 sensors installed in blocks 9 (TZM core) and 14 (tungsten core) at different intensities.  

The temperature measurements presented were taken at a given time frame where it is considered that steady-state regime had been reached and that the temperature measured is close to the average value of temperature taking into account the oscillations of $\pm\SI{0.4}{\celsius}$ in the supply temperature (as presented in Figure~\ref{fig:TGT:proto_temp_vs_int}). The Pt100 measurement accuracy displayed in Figure~\ref{fig:proto_9_evo} has been calculated from the tolerance of the Pt100 which is $\pm(0,3 + 0,005*T)^{\circ}$C for a given temperature $T$ (class B accuracy).  

As shown in Figure~\ref{fig:proto_9_evo}, the measured temperature evolution shows a very similar trend to the one calculated by means of FEM simulations. It can also be observed that the comparison between the measured and calculated temperatures presents larger deviations at high intensity than for the first pulses at low intensity. This can be justified by the fact that the high intensity pulses took place after several hours of operation: under these conditions, the initial parameters necessary for the FEM and CFD calculations are more difficult to estimate. Different initial conditions could modify the flow behaviour and, thus, the heat dissipation from the blocks. The cool-down phase after beam impact is more affected by these potential changes, explaining the bigger discrepancy between the measured and calculated values in that period. In any case, the comparison between the measured and calculated values is still within a relative deviation of less than 10\% as described in Figure~\ref{fig:TGT:proto_temp_vs_int}.

The temperature sensors are placed at a distance of 20 mm radially from the beam axis. Therefore, the temperatures measured by the Pt100 sensors are presumably lower than the maximum values reached during the target prototype operation. Taking into account the measured values of temperature during the test, and assuming a good correlation with the thermal simulations (within an acceptable deviation of about 10\%), the maximum temperatures found in the target core can be estimated. For an intensity of 3.75$\cdot10^{12}$ ppp, which is the maximum reached during the first day of tests, the maximum temperatures expected in the core and cladding materials of the target prototype are presented in Table~\ref{tab:TGT:proto_test_temps}, compared with the operational temperatures foreseen in the final BDF target~\cite{LopezSola:2019sfp}. The temperature levels attained in the core and cladding materials during the target prototype testing are higher than the ones foreseen in the final target under operational conditions.

\begin{table*}[hbtp]
\begin{ruledtabular}
\caption{\label{tab:TGT:proto_test_temps} Maximum temperatures measured by the sensors (T$_{measured}$) and corresponding expected temperature according to the FEM simulations (T$_{sim}$). Pt100 accuracy computed as $\pm(0,3 + 0,005*T)^{\circ}$C. Estimation of the maximum temperatures reached in the different target materials at 3.75$\cdot10^{12}$ ppp (T$_{max,sim}$) and comparison with the maximum temperatures foreseen for the final target under operational conditions. The target block where the maximum temperature is measured/expected is shown in parenthesis. The current design of the BDF final target does not include pure tantalum as cladding material.}
\begin{tabular}{lllll}
\multirow{3}{*}{Material} & \multicolumn{3}{c}{Prototype target} & Final target \\ \cline{2-4}
 & T$_{measured}$ &  T$_{sim}$ & T$_{max,sim}$ & T$_{max,sim}$ \\
 \hline
TZM & - & - & \SI{280}{\celsius} (4) & \SI{180}{\celsius} (9) \\
W & - & - & \SI{160}{\celsius} (14) & \SI{150}{\celsius} (14) \\
Ta2.5W & 38.8$\pm\SI{0.5}{\celsius}$ (9) & \SI{40}{\celsius} (9) & \SI{250}{\celsius} (4) & \SI{160}{\celsius} (4) \\
Ta & 46$\pm\SI{0.5}{\celsius}$ (14)$^{*}$ & \SI{43.8}{\celsius} (14) & \SI{195}{\celsius} (8) & - 
\end{tabular}
\end{ruledtabular}
\vspace{-0.8em}
 \begin{flushleft}
\footnotesize{$^{*}$\SI{60}{\celsius} were measured by the upper Pt100 of block 8, but this measurement is considered non-representative.}
\end{flushleft}
\end{table*}

\subsection{Strain measurements and stress evaluation}
\label{Sec:TGT:Proto:tests:strain}

\begin{figure}
\centering %
\includegraphics[width=0.45\textwidth]{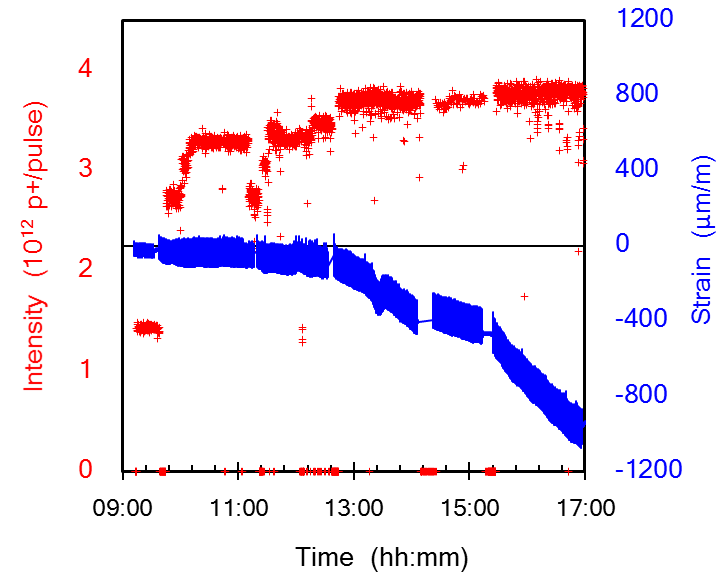}
\caption{\label{fig:proto_strain_drift} Transversal strain measured by one of the strain gauges placed in block 9 (T\_01) throughout the execution of the tests, and beam intensity on target during the same period of time. A drift in the strain values measured can be clearly noticed, with a higher impact during high intensity beam periods.}
\end{figure}

Radial and transversal strain measurements were performed at three different points of each instrumented block (see Figure~\ref{fig:TGT:sensors_position}). During the execution of the tests, a "drift" in the values measured was observed in all the strain gauges, as illustrated by Figure~\ref{fig:proto_strain_drift}. It can be seen that this effect is specially marked during high intensity and high repetition rate beam operation. The measurements deviation was usually observed towards negative values as in Figure~\ref{fig:proto_strain_drift}, but was also found towards positive values in some of the gauges. 

This drift has been previously reported in literature~\cite{Radiation_instru}, and is related to radiation effects on the glue attaching the gauges to the blocks, given the high dose rate to which the instrumentation is exposed (see Section~\ref{Sec:TGT:Proto:Instru}). Other options such as radiation-resistant fiber optic strain sensors as employed at the Spallation Neutron Source at ORNL~\cite{Straingauge_SNS} could be explored in the future to avoid the aforementioned effect.

Most of the strain measurements used for the present analysis correspond to the relative values of strain (strain variation during beam impact), which are not affected by the radiation drift. The absolute measured strain is not considered to be representative, except for the very first pulses on target where a minimum impact of the drift is observed.

\subsubsection{Strain evolution - first beam pulses}

\begin{figure}
\centering %
\includegraphics[width=0.47\textwidth]{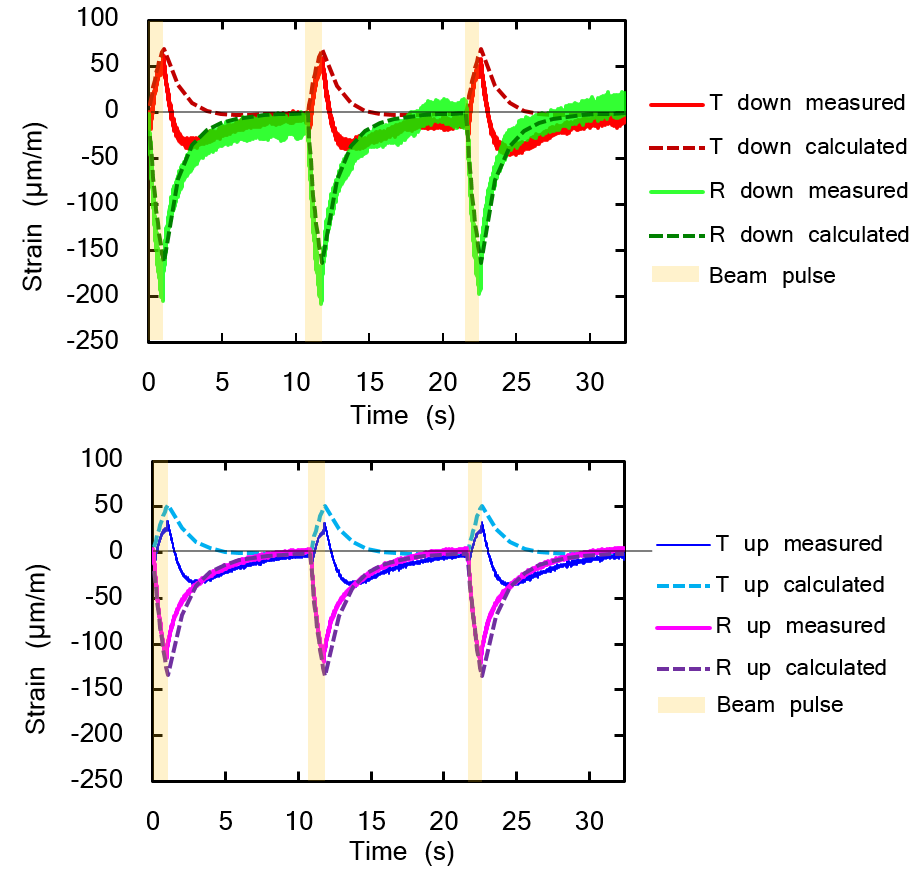}
\caption{\label{fig:proto_strain_evo} Radial (R) and transversal (T) strain measured by two strain gauges placed in positions 01 (down) and 03 (up) of block 4 (see Figure~\ref{fig:TGT:sensors_position}). Evolution during the first three consecutive beam pulses on target, at an intensity of 1.5$\cdot10^{12}$ ppp and a repetition rate of 10.8 seconds. Comparison with the evolution predicted by means of FEM simulations.}
\end{figure}

Figure~\ref{fig:proto_strain_evo} depicts the radial and transversal strain evolution measured by the strain gauges placed in block 4 during the first three beam pulses on target, at an intensity of 1.5$\cdot10^{12}$ ppp. It is considered that at this stage the drift due to radiation effects did not affect the measurements, therefore the absolute strain is seen as physically correct. The strain gauges in position 02 (R\_02 and T\_02) did not provide any physical measurement due to malfunction of the gauges. 

It can be observed that, during the beam impact, the transversal strain recorded is purely tensile and the radial strain is purely compressive, as predicted by the structural simulations. The strain variation measured by the strain gauges placed in the lower half of the block (T\_01 and R\_01) is higher than for the other gauges, due to their proximity to the beam impact position (around 2 mm down with respect to the target axis).

As depicted in Figure~\ref{fig:proto_strain_evo}, a good comparison is found between the trend of the measured values and the ones predicted by means of FEM calculations. The agreement between the measured and calculated values is less accurate in the cool-down phase after beam impact, in particular for the transversal strain. 

This difference could be explained by several factors, such as the precise way in which the target blocks are constrained by the inner tank, the asymmetry and exact location of the beam for a given pulse, the effect of the strain gauges' glue and protective cover, and the possible deformation of the strain gauge itself due to its interaction with the particle shower generated during the beam impact. Regarding the effect of the interaction between the secondary particles and the strain gauges, it shall be noted that the gauges are expected to present a similar deformation in both the radial and transversal directions due to this interaction. In favour of this hypothesis, it can be seen that the trend of the last 5-10 seconds of cool-down is almost identical for the radial and transversal strain measurements, which is observed in almost all the strain gauges. This issue will be further investigated in the future.

\subsubsection{Strain variation per pulse at different intensities}

\begin{figure*}
\centering %
\includegraphics[width=1\textwidth]{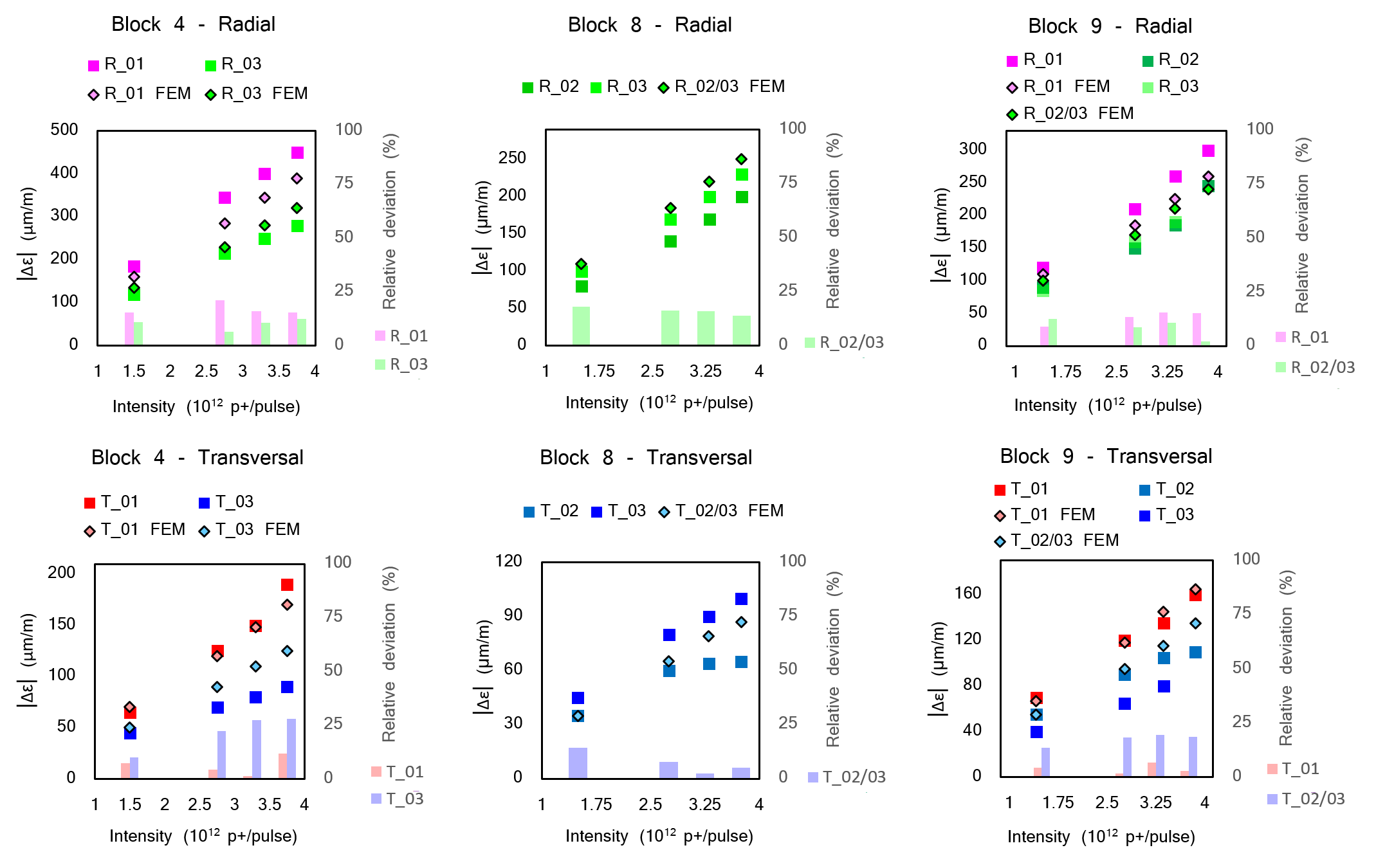}
\caption{\label{fig:TGT:proto_strain_vs_int} Strain variation measured in three of the instrumented blocks, in the transversal (T) and radial (R) directions for positions 01 (down), 02 (up left) and 03 (up right), see Figure~\ref{fig:TGT:sensors_position}. The values represented correspond to the average value of strain variation ($|\Delta\varepsilon|$) measured at different intensities once steady-state regime was reached: 1.5, 2.75, 3.25$\cdot10^{12}$ ppp (10.8 seconds cycle) and 3.75$\cdot10^{12}$ ppp (7.2 seconds cycle). A comparison with the strain variation obtained from the FEM simulations is shown, as well as the relative deviation between the measured and calculated values. Note that the measurements of some of the gauges have not been plotted due to the lack of signal.}
\end{figure*}

Figure~\ref{fig:TGT:proto_strain_vs_int} presents the variation of transversal and radial strain after one pulse at different intensities for three of the instrumented blocks. The values reported correspond to the average value of strain variation per pulse (which will be simply referred to as ``strain variation'' in what follows), $|\Delta\varepsilon|$, measured once steady-state regime was reached under a given repetition rate. The modulus of the strain variation is plotted since this value is positive in the transversal direction during the beam impact (tensile strain) and negative in the radial direction (compressive strain), as shown in Figure~\ref{fig:proto_strain_evo}. The strain variation calculated through thermo-structural calculations is also displayed in the figure for comparison. 

The strain variation during beam impact increases for greater values of intensity, as predicted by the FEM calculations. Due to the beam offset towards the bottom of the target axis, the strain variation measured in the strain gauge position 01 is always larger than in positions 02 and 03. As expected, the strain variation measured in positions 02 and 03 is quite close in most of the cases (a clear measurement in both positions was provided only in blocks 8 and 9). The difference between the measurements in positions 02 and 03 is thought to be caused by slight asymmetries of the beam in the horizontal plane. Given that in the thermal calculations the beam was assumed to be symmetric, the same strain values are obtained in the FEM simulations for positions 02 and 03; the deviation between the measured and calculated values was calculated by comparing the average of the measurement in both positions with the FEM-calculated one. 

\interfootnotelinepenalty=10000

As shown in Figure~\ref{fig:TGT:proto_strain_vs_int}, the relative deviation between the measured and expected values is below 25\%, which is considered to be an acceptable deviation given the large number of uncertainties that could influence the analysis results\footnote{First, the thermal calculations that are used as an input for the structural simulations are affected by a series of factors (described in the previous section) that could introduce discrepancies between the temperature measurements and the simulation results. Then, additional elements could alter the strain measurements, such as the way in which the blocks are constrained by the inner tank, the location and size of the beam, as well as the effect of the glue and the protective cover of the gauges.}. 

An exceptional case is found in block 14, where the measurements recorded at low intensities present a large deviation with respect to the calculated values: the increase of transversal strain is two to three times higher than expected, and the decrease of radial strain is less than half of the predicted value. Only in position 02 was the measured transversal strain coherent with the FEM values for all the intensities. Moreover, the radial strain variation measured in positions 01 and 03 (R\_01 and R\_03) was fundamentally tensile instead of compressive. It is expected that the PIE results will shed light on this issue, which will need further investigation. 

At high intensity, 3.75$\cdot10^{12}$ ppp, the average strain variation measured by all the gauges in block 14 (except for R\_01 and R\_03) was consistent with the predicted values, within a relative deviation of less than 10\%. Table~\ref{tab:TGT:proto_strain_14} summarizes the average strain variation measured at 3.75$\cdot10^{12}$ ppp as well as a comparison with the FEM simulation results and the subsequent deviation.

\begin{table}
\begin{ruledtabular}
\caption{\label{tab:TGT:proto_strain_14} The table presents the average strain variation per pulse measured at 3.75$\cdot10^{12}$ ppp in four of the strain gauges of block 14 in the radial and transversal directions. The expected strain variation obtained through FEM calculations and the relative deviations between both values are also displayed.}
\begin{tabular}{ccccc}
\multirow{2}{*}{Block} & \multirow{2}{*}{Strain gauge} & \multicolumn{2}{c}{Strain variation [\textmu m/m]} & Relative \\
& & Measured & Expected & deviation \\ \hline
\multirow{4}{*}{14} & T\_01 & 80 & 75 & 6.7 \% \\
 &T\_02 & 60 & 65 & 7.7 \% \\
 &T\_03 & 70 & 65 & 7.7 \% \\
 &R\_01 & 100$^{*}$ & -180 & - \\
 &R\_02 & -165 & -160 & 3.1 \% \\
 &R\_03 & 110$^{*}$ & -160 & -
\end{tabular}
\end{ruledtabular}
\vspace{-0.8em}
 \begin{flushleft}
\footnotesize{$^{*}$The radial strain variation measured in positions 01 and 03 was tensile instead of compressive and therefore the relative deviation is considered non-representative. Further analysis during the PIE phase are required to clarify the origin of this discrepancy.}
\end{flushleft}
\end{table}

\subsubsection{Estimation of maximum stress and stress amplitude}

According to the structural simulations, the state of stresses in the measuring points is expected to be mainly bi-axial, the transversal and radial directions being the two principal directions of the stresses. Therefore, the measurements of strain variation in the radial and transversal directions ($\Delta\varepsilon_R$ and $\Delta\varepsilon_T$) can be used to calculate the stress variation per pulse in the principal directions ($\Delta\sigma_R$ and $\Delta\sigma_T$), according to the following relation:

\begin{equation}
\label{{eq:stress_variation}}
\begin{aligned}
    \Delta\sigma_R = E/(1-\nu^2) \cdot (\Delta\varepsilon_R + \nu\cdot\Delta\varepsilon_T)\ \ \ ,\\
    \Delta\sigma_T = E/(1-\nu^2) \cdot (\Delta\varepsilon_T + \nu\cdot\Delta\varepsilon_R)\ \ \ \ 
\end{aligned}
\end{equation}

where E is the Young's modulus and $\nu$ the Poisson's ratio of the material (pure tantalum or Ta2.5W). 

In terms of fatigue life, the aforementioned stress variation $\Delta\sigma$ corresponds to the stress range in the principal directions; it can be used to calculate the stress amplitude $\sigma_a$ following $\sigma_a = \Delta\sigma/2$. From the stress amplitude in the principal directions, an equivalent stress amplitude has been calculated, $\sigma_{a,eq}$, that is expected to give the same fatigue life in uni-axial loading as the bi-axial stress-state found in the measuring points. The octahedral shear stress (von Mises) theory has been used for that purpose, given that tantalum and Ta2.5W are ductile materials and that the principal directions remain unchanged during the loading cycle (proportional loading). This equivalent stress amplitude obtained from the measured values of strain can be compared to the equivalent stress amplitude expected in the measuring points from the FEM simulations.

\begin{table*}
\begin{ruledtabular}
\caption{\label{tab:TGT:proto_test_strain} The table presents the maximum average strain variation per pulse ($\Delta\varepsilon$) in the transversal (T) and radial (R) directions measured during the BDF target prototype beam tests in the two cladding materials (Ta and Ta2.5W). The values were measured in blocks 4 (Ta2.5W) and 8 (Ta) under a high intensity beam (3.75$\cdot10^{12}$ ppp) after steady-state regime was reached. The radial and transversal stress variation ($\Delta\sigma$), and equivalent stress amplitude ($\sigma_{a,eq}$) calculated in the measuring points is also displayed, as well as a comparison with the expected stress amplitude calculated by means of FEM simulations. The table also shows the maximum stress amplitude and the maximum von Mises equivalent stress reached in the target prototype materials during the beam tests under high intensity operation, estimated through thermo-mechanical simulations, and for comparison the equivalent stress amplitude and von Mises equivalent stress foreseen during the final BDF target operation are also presented~\cite{LopezSola:2019sfp}.}
\begin{tabular}{ccccccccccc}
\multirow{4}{*}{\begin{tabular}[c]{@{}c@{}}Material\\(block no.)
\end{tabular}} & \multicolumn{8}{c}{\hspace{-2.5em}Target prototype} & \multicolumn{2}{c}{Final target} \\
 & \multicolumn{5}{c}{Strain gauge measurements} & \multicolumn{3}{c}{FEM calculations} & \multicolumn{2}{c}{FEM calculations} \\ \cline{2-6} \cline{7-9}  \cline{10-11} 
 & \multicolumn{2}{c}{\begin{tabular}[c]{@{}c@{}}$\Delta\varepsilon$ \\ (\textmu m/m) \end{tabular}} &
 \multicolumn{2}{c}{\begin{tabular}[c]{@{}c@{}}$\Delta\sigma$\\ (MPa)\end{tabular}} & \multirow{2}{*}{\begin{tabular}[c]{@{}c@{}}$\sigma_{a,eq}$\\(MPa)\end{tabular}} & {\begin{tabular}[c]{@{}c@{}}Expected \\$\sigma_{a,eq}$\end{tabular}} & {\begin{tabular}[c]{@{}c@{}}Max. estimated \\$\sigma_{a,eq}$\end{tabular}} & {\begin{tabular}[c]{@{}c@{}}Max. estimated\\von Mises eq.\end{tabular}} & \multirow{2}{*}{\begin{tabular}[c]{@{}c@{}}Max. foreseen\\$\sigma_{a,eq}$ (MPa)\end{tabular}} & {\begin{tabular}[c]{@{}c@{}}Max. foreseen\\von Mises eq.\end{tabular}} \\
 & T & R & T & R &  &  (MPa) & (MPa) & stress (MPa) &  & stress (MPa) \\ \hline
TZM (4) & - & - & - & - & - & - & 80   & 180   & 60 & 130 \\
W (14) & - & - & - & - & - & - & 35  & 115   & 30 & 95 \\
Ta2.5W (4) & 190 & -450 & 13 & -80 & 43 & 37 & 50   & 105  & 45 & 95 \\
Ta (8) & 100 & -230 & 7 & -40 & 22 & 23 & 30   & 75   & - & -
\end{tabular}
\end{ruledtabular}
\end{table*}

Moreover, assuming a good correlation between the measurements and the thermo-mechanical simulations, within a 25\% margin as shown in Figure~\ref{fig:TGT:proto_strain_vs_int}, the maximum value of stress amplitude reached in the target materials (presumably reached closer to the beam axis) has been estimated from the FEM calculations. The maximum von Mises equivalent stress expected during the beam tests has also been obtained. Finally, those values have been compared with the stress amplitude and equivalent von Mises stress foreseen in the final BDF target operation~\cite{LopezSola:2019sfp}. Table~\ref{tab:TGT:proto_test_strain} presents a summary of the aforementioned values for each one of the target materials. 

Several conclusions can be drawn from the results presented in Table~\ref{tab:TGT:proto_test_strain}. First, it can be seen that there is a good correlation between the measured and the expected values of equivalent stress amplitude in the measuring points, with a deviation of less than 15\% for Ta2.5W and Ta. Then, assuming that the results of the FEM simulations are coherent with the behaviour of the target prototype during the tests (within the 25\% margin mentioned in the current Section), it can be seen that the target prototype was subjected to more challenging conditions in terms of structural loads with respect to those foreseen for the final target. From the material failure point of view, the maximum von Mises equivalent stress that was reached in the target prototype materials during the impact of a high intensity beam is higher than the one expected in the final target materials under normal operational conditions. 

From the fatigue life point of view, the equivalent stress amplitude to which the target materials were subjected during the target prototype tests is higher than the equivalent stress amplitude foreseen in the final target materials. However, it is worth recalling that the number of cycles reached during the whole duration of the prototype tests (N~$=10^4$) is lower than the number of cycles expected for the final BDF target operation (N~$=10^7$). It must be considered as well that for an accurate estimation of the fatigue life of the materials, the influence of the mean stress should also be taken into account. This parameter could not be determined from the strain gauges measurements, since the absolute values of strain were not considered as representative (see Figure~\ref{fig:proto_strain_drift}). 

Nevertheless, the prototype tests have provided an important cross-check of the target materials behaviour under high stresses and high-cycle fatigue, as well as a preliminary insight into the robustness of the core/cladding interface. A further assessment of the materials and bonding interfaces performance under the challenging conditions of the beam tests will be given during the Post Irradiation Examination of the target blocks (see Section~\ref{Sec:TGT:Proto:PIE}).

\section{Post Irradiation Experiment (PIE) plans}
\label{Sec:TGT:Proto:PIE}

A Post Irradiation Examination (PIE) campaign is foreseen for some of the irradiated blocks of the BDF target prototype, aiming at a better understanding of the target materials' response to the BDF target conditions. More precisely, the PIE has several objectives:

\begin{itemize}
    \item Study the state and integrity of the cladding surface to validate the resistance of the cladding materials to high-speed water cooling conditions. 
    \item Validate Ta2.5W as new cladding material and compare its performance with unalloyed Ta.
    \item Study the core/cladding interface to evaluate its reliability under thermal cycling due to the beam interaction with the target. Identify any possible degradation of the interface properties (strength or thermal conductivity) or presence of defects (detachments or segregations) for each different core/cladding couple (TZM-Ta2.5W, TZM-Ta and W-Ta)
    \item Similarly, study the core and cladding materials (W, TZM, Ta and Ta2.5W) to assess their performance after thermal cycling. Identify any possible degradation of the material properties (strength or thermal conductivity) or presence of defects (cracks, voids, microstructural changes or segregations)
    \item Check for any other potential concerns related to the target operational conditions (such as target blocks movement or deformation, cooling channels blocking, instrumentation damage, etc.), in order to provide representative feedback for the final target design.
\end{itemize}

The PIE is foreseen to be performed on several target blocks, representative of the different block dimensions, the different core and cladding material combinations, and the highest levels of cyclic stress and temperature. The selected blocks are the four instrumented blocks, whose characteristics are detailed in Table~\ref{tab:TGT:instrublocks}, plus block 3 (same dimensions and materials as block 4) and block 15 (same materials as block 14 but 80 mm length).

The PIE activities are foreseen to start in 2020, and will provide a first confirmation of the survivability of the final target under the BDF conditions, assessing the final target design in terms of material selection, mechanical design and manufacturing process, in a similar way of what has been done in Ref.~\cite{PhysRevAccelBeams.22.041001,MCCLINTOCK2014130}. The PIE will also help elucidating some of the discrepancies found between the target prototype simulations and experimental data.

\section{Conclusions}

A reduced scale prototype of the target for the future Beam Dump Facility has been designed, manufactured and tested under the high intensity SPS proton beam during 2018. The main objective of the beam tests was to provide a first validation of the BDF target design, as discussed in Ref.~\cite{LopezSola:2019sfp}.

The target prototype design and manufacture allowed for the identification of several challenges associated to the feasibility of the final target mechanical design, such as the diffusion bonding of clad refractory metals by means of HIP process, the production of a leak-tight target assembly with two concentric tanks compatible with an efficient water-cooling system. The experimental setup was located in the North Area of CERN, and required the installation of additional equipment such as beam instrumentation, protective shielding, and mechanical interfaces fully compatible with remote handling, all these elements being necessary for the safe and successful performance of the beam tests. 

The target prototype was tested under an ad-hoc tuned proton beam, whose characteristics were selected based on FEM thermal and structural calculations with the aim of achieving a meaningful reproduction of the final BDF target conditions despite the absence of dilution in the target prototype. The prototype cooling system design includes the main features of the final target cooling system; the pressure measurements carried out during operation have shown a good agreement with the CFD calculations and the high-speed and high-pressure water cooling circuit performed successfully during the tests. Most of the strain and temperature sensors installed in four blocks of the target prototype worked well throughout the tests. The good functioning of the instrumentation is considered to be an important achievement of the prototype tests, given the unprecedented conditions imposed by the testing environment, in particular regarding the high radiation dose and the high water speed and pressure to which the sensors are subjected. 

A first analysis of the online measurements has been carried out, showing that the values of temperature and bi-axial strain recorded are coherent with the predictions from the FEM calculations, within a relative deviation of 10\% and 25\% respectively. A larger deviation has been found in the strain measurements, due to the greater number of uncertainties in the structural calculations. As an outcome of this first analysis, it has been concluded the levels of temperatures and stresses of the final BDF target have been reached and even exceeded during the beam tests at high intensity. It has been estimated that the Ta2.5W cladding of the target prototype reached temperatures of \SI{250}{\celsius} and cyclic stresses of around 100 MPa.

The PIE campaign that will be carried out on some of the irradiated target prototype blocks will provide crucial information about the behaviour of the target materials under operational conditions similar to the ones of the future facility. Several activities are foreseen, including non-destructive and destructive testing on the bulk target materials and the core/cladding interface. 

In summary, the target prototype testing under the 400 GeV/c, 40 kW SPS primary beam has been successful, and is considered to be an important milestone in the validation of the future BDF target design.

\bibliography{references}

\begin{thebibliography}{27}%
\makeatletter
\providecommand \@ifxundefined [1]{%
 \@ifx{#1\undefined}
}%
\providecommand \@ifnum [1]{%
 \ifnum #1\expandafter \@firstoftwo
 \else \expandafter \@secondoftwo
 \fi
}%
\providecommand \@ifx [1]{%
 \ifx #1\expandafter \@firstoftwo
 \else \expandafter \@secondoftwo
 \fi
}%
\providecommand \natexlab [1]{#1}%
\providecommand \enquote  [1]{``#1''}%
\providecommand \bibnamefont  [1]{#1}%
\providecommand \bibfnamefont [1]{#1}%
\providecommand \citenamefont [1]{#1}%
\providecommand \href@noop [0]{\@secondoftwo}%
\providecommand \href [0]{\begingroup \@sanitize@url \@href}%
\providecommand \@href[1]{\@@startlink{#1}\@@href}%
\providecommand \@@href[1]{\endgroup#1\@@endlink}%
\providecommand \@sanitize@url [0]{\catcode `\\12\catcode `\$12\catcode
  `\&12\catcode `\#12\catcode `\^12\catcode `\_12\catcode `\%12\relax}%
\providecommand \@@startlink[1]{}%
\providecommand \@@endlink[0]{}%
\providecommand \url  [0]{\begingroup\@sanitize@url \@url }%
\providecommand \@url [1]{\endgroup\@href {#1}{\urlprefix }}%
\providecommand \urlprefix  [0]{URL }%
\providecommand \Eprint [0]{\href }%
\providecommand \doibase [0]{http://dx.doi.org/}%
\providecommand \selectlanguage [0]{\@gobble}%
\providecommand \bibinfo  [0]{\@secondoftwo}%
\providecommand \bibfield  [0]{\@secondoftwo}%
\providecommand \translation [1]{[#1]}%
\providecommand \BibitemOpen [0]{}%
\providecommand \bibitemStop [0]{}%
\providecommand \bibitemNoStop [0]{.\EOS\space}%
\providecommand \EOS [0]{\spacefactor3000\relax}%
\providecommand \BibitemShut  [1]{\csname bibitem#1\endcsname}%
\let\auto@bib@innerbib\@empty
\bibitem [{\citenamefont {Bonivento}\ \emph {et~al.}(2013)\citenamefont
  {Bonivento} \emph {et~al.}}]{EOI_SHiP}%
  \BibitemOpen
  \bibfield  {author} {\bibinfo {author} {\bibfnamefont {W.}~\bibnamefont
  {Bonivento}} \emph {et~al.},\ }\bibfield  {title} {\enquote {\bibinfo {title}
  {{Proposal to Search for Heavy Neutral Leptons at the SPS}},}\ }\href@noop {}
  {\  (\bibinfo {year} {2013})},\ \Eprint {http://arxiv.org/abs/1310.1762}
  {arXiv:1310.1762 [hep-ex]} \BibitemShut {NoStop}%
\bibitem [{\citenamefont {{S. Alekhin \textit{et al.} (SHiP
  Collaboration)}}(2016)}]{Alekhin_SHiP}%
  \BibitemOpen
  \bibfield  {author} {\bibinfo {author} {\bibnamefont {{S. Alekhin \textit{et
  al.} (SHiP Collaboration)}}},\ }\bibfield  {title} {\enquote {\bibinfo
  {title} {{A facility to Search for Hidden Particles at the CERN SPS: the SHiP
  physics case}},}\ }\href {\doibase 10.1088/0034-4885/79/12/124201} {\bibfield
   {journal} {\bibinfo  {journal} {Rept. Prog. Phys.}\ }\textbf {\bibinfo
  {volume} {79}},\ \bibinfo {pages} {124201} (\bibinfo {year} {2016})},\
  \Eprint {http://arxiv.org/abs/1504.04855} {arXiv:1504.04855 [hep-ph]}
  \BibitemShut {NoStop}%
\bibitem [{\citenamefont {Anelli}\ \emph {et~al.}(2015)\citenamefont {Anelli}
  \emph {et~al.}}]{Anelli_SHiP}%
  \BibitemOpen
  \bibfield  {author} {\bibinfo {author} {\bibfnamefont {M.}~\bibnamefont
  {Anelli}} \emph {et~al.} (\bibinfo {collaboration} {SHiP Collaboration}),\
  }\bibfield  {title} {\enquote {\bibinfo {title} {{A facility to Search for
  Hidden Particles (SHiP) at the CERN SPS}},}\ }\href@noop {} {\  (\bibinfo
  {year} {2015})},\ \Eprint {http://arxiv.org/abs/1504.04956} {arXiv:1504.04956
  [physics.ins-det]} \BibitemShut {NoStop}%
\bibitem [{\citenamefont {\textit{et al.}}(2019{\natexlab{a}})}]{Ahdida_2019}%
  \BibitemOpen
  \bibfield  {author} {\bibinfo {author} {\bibfnamefont {C.~Ahdida}\
  \bibnamefont {\textit{et al.}}},\ }\bibfield  {title} {\enquote {\bibinfo
  {title} {The experimental facility for the search for hidden particles at the
  {CERN} {SPS}},}\ }\href {\doibase 10.1088/1748-0221/14/03/p03025} {\bibfield
  {journal} {\bibinfo  {journal} {Journal of Instrumentation}\ }\textbf
  {\bibinfo {volume} {14}},\ \bibinfo {pages} {P03025--P03025} (\bibinfo {year}
  {2019}{\natexlab{a}})}\BibitemShut {NoStop}%
\bibitem [{\citenamefont {Lopez~Sola}\ \emph {et~al.}(2019)\citenamefont
  {Lopez~Sola} \emph {et~al.}}]{LopezSola:2019sfp}%
  \BibitemOpen
  \bibfield  {author} {\bibinfo {author} {\bibfnamefont {E.}~\bibnamefont
  {Lopez~Sola}} \emph {et~al.},\ }\bibfield  {title} {\enquote {\bibinfo
  {title} {{Design of a high power production target for the Beam Dump Facility
  at CERN}},}\ }\href@noop {} {\  (\bibinfo {year} {2019})},\ \Eprint
  {http://arxiv.org/abs/1904.03074} {arXiv:1904.03074 [physics.ins-det]}
  \BibitemShut {NoStop}%
\bibitem [{\citenamefont {\textit{et al.}}(2018{\natexlab{a}})}]{BDFcomplex}%
  \BibitemOpen
  \bibfield  {author} {\bibinfo {author} {\bibfnamefont {K.~Kershaw}\
  \bibnamefont {\textit{et al.}}},\ }\bibfield  {title} {\enquote {\bibinfo
  {title} {Design development for the {Beam Dump Facility Target Complex} at
  {CERN}},}\ }\href {\doibase 10.1088/1748-0221/13/10/p10011} {\bibfield
  {journal} {\bibinfo  {journal} {Journal of Instrumentation}\ }\textbf
  {\bibinfo {volume} {13}},\ \bibinfo {pages} {P10011--P10011} (\bibinfo {year}
  {2018}{\natexlab{a}})}\BibitemShut {NoStop}%
\bibitem [{\citenamefont {Gaillard}\ \emph {et~al.}(2011)\citenamefont
  {Gaillard} \emph {et~al.}}]{HRMT}%
  \BibitemOpen
  \bibfield  {author} {\bibinfo {author} {\bibfnamefont {H.}~\bibnamefont
  {Gaillard}} \emph {et~al.},\ }\bibfield  {title} {\enquote {\bibinfo {title}
  {{HiRadMat: A New Irradiation Facility for Material Testing at CERN}},}\
  }\bibfield  {booktitle} {\emph {\bibinfo {booktitle} {{Particle accelerator.
  Proceedings, 2nd International Conference, IPAC 2011, San Sebastian, Spain,
  September 4-9, 2011}}},\ }\href@noop {} {\bibfield  {journal} {\bibinfo
  {journal} {Conf. Proc.}\ }\textbf {\bibinfo {volume} {C110904}},\ \bibinfo
  {pages} {1665--1667} (\bibinfo {year} {2011})}\BibitemShut {NoStop}%
\bibitem [{\citenamefont {Torregrosa~Martin}\ \emph {et~al.}(2018)\citenamefont
  {Torregrosa~Martin}, \citenamefont {Calviani}, \citenamefont
  {Perillo-Marcone}, \citenamefont {Ferriere}, \citenamefont {Solieri},
  \citenamefont {Butcher}, \citenamefont {Grec},\ and\ \citenamefont
  {Canhoto~Espadanal}}]{PhysRevAccelBeams.21.073001}%
  \BibitemOpen
  \bibfield  {author} {\bibinfo {author} {\bibfnamefont {C.}~\bibnamefont
  {Torregrosa~Martin}}, \bibinfo {author} {\bibfnamefont {M.}~\bibnamefont
  {Calviani}}, \bibinfo {author} {\bibfnamefont {A.}~\bibnamefont
  {Perillo-Marcone}}, \bibinfo {author} {\bibfnamefont {R.}~\bibnamefont
  {Ferriere}}, \bibinfo {author} {\bibfnamefont {N.}~\bibnamefont {Solieri}},
  \bibinfo {author} {\bibfnamefont {M.}~\bibnamefont {Butcher}}, \bibinfo
  {author} {\bibfnamefont {L.-M.}\ \bibnamefont {Grec}}, \ and\ \bibinfo
  {author} {\bibfnamefont {J.}~\bibnamefont {Canhoto~Espadanal}},\ }\bibfield
  {title} {\enquote {\bibinfo {title} {{Scaled prototype of a tantalum target
  embedded in expanded graphite for antiproton production: Design,
  manufacturing, and testing under proton beam impacts}},}\ }\href {\doibase
  10.1103/PhysRevAccelBeams.21.073001} {\bibfield  {journal} {\bibinfo
  {journal} {Phys. Rev. Accel. Beams}\ }\textbf {\bibinfo {volume} {21}},\
  \bibinfo {pages} {073001} (\bibinfo {year} {2018})}\BibitemShut {NoStop}%
\bibitem [{\citenamefont {Torregrosa~Martin}\ \emph {et~al.}(2019)\citenamefont
  {Torregrosa~Martin}, \citenamefont {Perillo-Marcone}, \citenamefont
  {Calviani}, \citenamefont {Gentini}, \citenamefont {Butcher},\ and\
  \citenamefont {Mu\~noz Cobo}}]{PhysRevAccelBeams.22.013401}%
  \BibitemOpen
  \bibfield  {author} {\bibinfo {author} {\bibfnamefont {C.}~\bibnamefont
  {Torregrosa~Martin}}, \bibinfo {author} {\bibfnamefont {A.}~\bibnamefont
  {Perillo-Marcone}}, \bibinfo {author} {\bibfnamefont {M.}~\bibnamefont
  {Calviani}}, \bibinfo {author} {\bibfnamefont {L.}~\bibnamefont {Gentini}},
  \bibinfo {author} {\bibfnamefont {M.}~\bibnamefont {Butcher}}, \ and\
  \bibinfo {author} {\bibfnamefont {J.-L.}\ \bibnamefont {Mu\~noz Cobo}},\
  }\bibfield  {title} {\enquote {\bibinfo {title} {{Experiment exposing
  refractory metals to impacts of $440\text{ }\text{ }\mathrm{GeV}/c$ proton
  beams for the future design of the CERN antiproton production target:
  Experiment design and online results}},}\ }\href {\doibase
  10.1103/PhysRevAccelBeams.22.013401} {\bibfield  {journal} {\bibinfo
  {journal} {Phys. Rev. Accel. Beams}\ }\textbf {\bibinfo {volume} {22}},\
  \bibinfo {pages} {013401} (\bibinfo {year} {2019})}\BibitemShut {NoStop}%
\bibitem [{\citenamefont {\textit{et
  al.}}(2019{\natexlab{b}})}]{doi:10.1002/mdp2.33}%
  \BibitemOpen
  \bibfield  {author} {\bibinfo {author} {\bibfnamefont {F.-X.~Nuiry}\
  \bibnamefont {\textit{et al.}}},\ }\bibfield  {title} {\enquote {\bibinfo
  {title} {{3D Carbon/Carbon composites for beam intercepting devices at
  CERN}},}\ }\href {\doibase 10.1002/mdp2.33} {\bibfield  {journal} {\bibinfo
  {journal} {Material Design \& Processing Communications}\ }\textbf {\bibinfo
  {volume} {1}},\ \bibinfo {pages} {e33} (\bibinfo {year}
  {2019}{\natexlab{b}})},\ \bibinfo {note} {e33 MDPC-2018-013.R1}\BibitemShut
  {NoStop}%
\bibitem [{\citenamefont {Davenne}\ \emph {et~al.}(2018)\citenamefont
  {Davenne}, \citenamefont {Loveridge}, \citenamefont {Bingham}, \citenamefont
  {Wark}, \citenamefont {Back}, \citenamefont {Caretta}, \citenamefont
  {Densham}, \citenamefont {O'Dell}, \citenamefont {Wilcox},\ and\
  \citenamefont {Fitton}}]{Davenne:2018nxs}%
  \BibitemOpen
  \bibfield  {author} {\bibinfo {author} {\bibfnamefont {T.}~\bibnamefont
  {Davenne}}, \bibinfo {author} {\bibfnamefont {P.}~\bibnamefont {Loveridge}},
  \bibinfo {author} {\bibfnamefont {R.}~\bibnamefont {Bingham}}, \bibinfo
  {author} {\bibfnamefont {J.}~\bibnamefont {Wark}}, \bibinfo {author}
  {\bibfnamefont {J.~J.}\ \bibnamefont {Back}}, \bibinfo {author}
  {\bibfnamefont {O.}~\bibnamefont {Caretta}}, \bibinfo {author} {\bibfnamefont
  {C.}~\bibnamefont {Densham}}, \bibinfo {author} {\bibfnamefont
  {J.}~\bibnamefont {O'Dell}}, \bibinfo {author} {\bibfnamefont
  {D.}~\bibnamefont {Wilcox}}, \ and\ \bibinfo {author} {\bibfnamefont
  {M.}~\bibnamefont {Fitton}},\ }\bibfield  {title} {\enquote {\bibinfo {title}
  {{Observed proton beam induced disruption of a tungsten powder sample at
  CERN}},}\ }\href {\doibase 10.1103/PhysRevAccelBeams.21.073002} {\bibfield
  {journal} {\bibinfo  {journal} {Phys. Rev. Accel. Beams}\ }\textbf {\bibinfo
  {volume} {21}},\ \bibinfo {pages} {073002} (\bibinfo {year}
  {2018})}\BibitemShut {NoStop}%
\bibitem [{\citenamefont {\textit{et
  al.}}(2018{\natexlab{b}})}]{PhysRevAccelBeams.21.033401}%
  \BibitemOpen
  \bibfield  {author} {\bibinfo {author} {\bibfnamefont {O.~Caretta}\
  \bibnamefont {\textit{et al.}}},\ }\bibfield  {title} {\enquote {\bibinfo
  {title} {Proton beam induced dynamics of tungsten granules},}\ }\href
  {\doibase 10.1103/PhysRevAccelBeams.21.033401} {\bibfield  {journal}
  {\bibinfo  {journal} {Phys. Rev. Accel. Beams}\ }\textbf {\bibinfo {volume}
  {21}},\ \bibinfo {pages} {033401} (\bibinfo {year}
  {2018}{\natexlab{b}})}\BibitemShut {NoStop}%
\bibitem [{\citenamefont {\textit{et al.}}(2014)}]{PhysRevSTAB.17.021004}%
  \BibitemOpen
  \bibfield  {author} {\bibinfo {author} {\bibfnamefont {M.~Cauchi}\
  \bibnamefont {\textit{et al.}}},\ }\bibfield  {title} {\enquote {\bibinfo
  {title} {{High energy beam impact tests on a LHC tertiary collimator at the
  CERN high-radiation to materials facility}},}\ }\href {\doibase
  10.1103/PhysRevSTAB.17.021004} {\bibfield  {journal} {\bibinfo  {journal}
  {Phys. Rev. ST Accel. Beams}\ }\textbf {\bibinfo {volume} {17}},\ \bibinfo
  {pages} {021004} (\bibinfo {year} {2014})}\BibitemShut {NoStop}%
\bibitem [{\citenamefont {\textit{et al.}}(2007)}]{COMPASS}%
  \BibitemOpen
  \bibfield  {author} {\bibinfo {author} {\bibfnamefont {P.~Abbon}\
  \bibnamefont {\textit{et al.}}},\ }\bibfield  {title} {\enquote {\bibinfo
  {title} {{The COMPASS experiment at CERN}},}\ }\href {\doibase
  https://doi.org/10.1016/j.nima.2007.03.026} {\bibfield  {journal} {\bibinfo
  {journal} {Nuclear Instruments and Methods in Physics Research Section A:
  Accelerators, Spectrometers, Detectors and Associated Equipment}\ }\textbf
  {\bibinfo {volume} {577}},\ \bibinfo {pages} {455 -- 518} (\bibinfo {year}
  {2007})}\BibitemShut {NoStop}%
\bibitem [{\citenamefont {Bravin}\ \emph {et~al.}(2005)\citenamefont {Bravin},
  \citenamefont {Burger}, \citenamefont {Ferioli}, \citenamefont {Focker},
  \citenamefont {Guerrero},\ and\ \citenamefont {Maccaferri}}]{Bravin:2005vb}%
  \BibitemOpen
  \bibfield  {author} {\bibinfo {author} {\bibfnamefont {E.}~\bibnamefont
  {Bravin}}, \bibinfo {author} {\bibfnamefont {S.}~\bibnamefont {Burger}},
  \bibinfo {author} {\bibfnamefont {G.}~\bibnamefont {Ferioli}}, \bibinfo
  {author} {\bibfnamefont {G.~J.}\ \bibnamefont {Focker}}, \bibinfo {author}
  {\bibfnamefont {A.}~\bibnamefont {Guerrero}}, \ and\ \bibinfo {author}
  {\bibfnamefont {R.}~\bibnamefont {Maccaferri}},\ }\bibfield  {title}
  {\enquote {\bibinfo {title} {{A new TV beam observation system for CERN}},}\
  }in\ \href {http://weblib.cern.ch/abstract?CERN-AB-2005-076} {\emph {\bibinfo
  {booktitle} {{7th European Workshop on Beam Diagnostics and Instrumentation
  for Particle Accelerators (DIPAC 2005) Lyon, France, June 6-8, 2005}}}}\
  (\bibinfo {year} {2005})\ pp.\ \bibinfo {pages} {212--214}\BibitemShut
  {NoStop}%
\bibitem [{\citenamefont {{A. T. Nelson, J. A. O'Toole, R. A. Valicenti, S. A.
  Maloy}}(2012)}]{HIP1}%
  \BibitemOpen
  \bibfield  {author} {\bibinfo {author} {\bibnamefont {{A. T. Nelson, J. A.
  O'Toole, R. A. Valicenti, S. A. Maloy}}},\ }\bibfield  {title} {\enquote
  {\bibinfo {title} {{Fabrication of a tantalum-clad tungsten target for
  LANSCE}},}\ }\href {\doibase 10.1016/j.jnucmat.2011.11.041} {\bibfield
  {journal} {\bibinfo  {journal} {J. Nucl. Mater.}\ }\textbf {\bibinfo {volume}
  {431}},\ \bibinfo {pages} {172--184} (\bibinfo {year} {2012})}\BibitemShut
  {NoStop}%
\bibitem [{\citenamefont {{M. Kawai, K. Kikuchi, H. Kurishita, J.-F. Li, M.
  Furusaka}}(2001)}]{HIP2}%
  \BibitemOpen
  \bibfield  {author} {\bibinfo {author} {\bibnamefont {{M. Kawai, K. Kikuchi,
  H. Kurishita, J.-F. Li, M. Furusaka}}},\ }\bibfield  {title} {\enquote
  {\bibinfo {title} {{Fabrication of a tantalum-clad tungsten target for
  KENS}},}\ }\href@noop {} {\bibfield  {journal} {\bibinfo  {journal} {J. Nucl.
  Mater.}\ }\textbf {\bibinfo {volume} {296}},\ \bibinfo {pages} {312--320}
  (\bibinfo {year} {2001})}\BibitemShut {NoStop}%
\bibitem [{\citenamefont {Jones}\ and\ \citenamefont
  {Wilcox}(2018)}]{ISISclad}%
  \BibitemOpen
  \bibfield  {author} {\bibinfo {author} {\bibfnamefont {L.~G.}\ \bibnamefont
  {Jones}}\ and\ \bibinfo {author} {\bibfnamefont {D.}~\bibnamefont {Wilcox}},\
  }\bibfield  {title} {\enquote {\bibinfo {title} {{ISIS} {TS}1 project target
  {\textendash} design for manufacture},}\ }\href {\doibase
  10.1088/1742-6596/1021/1/012056} {\bibfield  {journal} {\bibinfo  {journal}
  {J. Phys.: Conference Series}\ }\textbf {\bibinfo {volume} {1021}},\ \bibinfo
  {pages} {012056} (\bibinfo {year} {2018})}\BibitemShut {NoStop}%
\bibitem [{\citenamefont {Busom~Descarrega}\ \emph {et~al.}(2019)\citenamefont
  {Busom~Descarrega}, \citenamefont {Calviani}, \citenamefont {Hutsch},
  \citenamefont {Lopez~Sola}, \citenamefont {Perez~Fontenla}, \citenamefont
  {Perillo-Marcone}, \citenamefont {Sgobba},\ and\ \citenamefont
  {Weißgaerber}}]{HIP_Busom}%
  \BibitemOpen
  \bibfield  {author} {\bibinfo {author} {\bibfnamefont {J.}~\bibnamefont
  {Busom~Descarrega}}, \bibinfo {author} {\bibfnamefont {M.}~\bibnamefont
  {Calviani}}, \bibinfo {author} {\bibfnamefont {T.}~\bibnamefont {Hutsch}},
  \bibinfo {author} {\bibfnamefont {E.}~\bibnamefont {Lopez~Sola}}, \bibinfo
  {author} {\bibfnamefont {A.~T.}\ \bibnamefont {Perez~Fontenla}}, \bibinfo
  {author} {\bibfnamefont {A.}~\bibnamefont {Perillo-Marcone}}, \bibinfo
  {author} {\bibfnamefont {S.}~\bibnamefont {Sgobba}}, \ and\ \bibinfo {author}
  {\bibfnamefont {T.}~\bibnamefont {Weißgaerber}},\ }\bibfield  {title}
  {\enquote {\bibinfo {title} {{Application of hot isostatic pressing (HIP)
  technology to diffusion bond refractory metals for proton beam targets and
  absorbers at CERN}},}\ }\href {\doibase 10.1002/mdp2.101} {\bibfield
  {journal} {\bibinfo  {journal} {{Material Design \& Processing
  Communications}}\ }\textbf {\bibinfo {volume} {0}},\ \bibinfo {pages} {e101}
  (\bibinfo {year} {2019})},\ \bibinfo {note} {e101 MDPC-2019-050}\BibitemShut
  {NoStop}%
\bibitem [{\citenamefont {CERN}(2018)}]{CDS_pictures1}%
  \BibitemOpen
  \bibfield  {author} {\bibinfo {author} {\bibnamefont {CERN}},\ }\bibfield
  {title} {\enquote {\bibinfo {title} {\text{BDF} target prototype assembly},}\
  }\href {http://cds.cern.ch/record/2635418} {\bibfield  {journal} {\bibinfo
  {journal} {CERN-PHOTO-201808-199}\ } (\bibinfo {year} {2018})}\BibitemShut
  {NoStop}%
\bibitem [{\citenamefont {{Weiler}}\ \emph {et~al.}(2007)\citenamefont
  {{Weiler}}, \citenamefont {{Aberle}}, \citenamefont {{Assmann}},
  \citenamefont {{Chamizo}}, \citenamefont {{Kadi}}, \citenamefont {{Lettry}},
  \citenamefont {{Losito}},\ and\ \citenamefont {{Redaelli}}}]{Collimator}%
  \BibitemOpen
  \bibfield  {author} {\bibinfo {author} {\bibfnamefont {T.}~\bibnamefont
  {{Weiler}}}, \bibinfo {author} {\bibfnamefont {O.}~\bibnamefont {{Aberle}}},
  \bibinfo {author} {\bibfnamefont {R.}~\bibnamefont {{Assmann}}}, \bibinfo
  {author} {\bibfnamefont {R.}~\bibnamefont {{Chamizo}}}, \bibinfo {author}
  {\bibfnamefont {Y.}~\bibnamefont {{Kadi}}}, \bibinfo {author} {\bibfnamefont
  {J.}~\bibnamefont {{Lettry}}}, \bibinfo {author} {\bibfnamefont
  {R.}~\bibnamefont {{Losito}}}, \ and\ \bibinfo {author} {\bibfnamefont
  {S.}~\bibnamefont {{Redaelli}}},\ }\bibfield  {title} {\enquote {\bibinfo
  {title} {{LHC collimation system hardware commissioning}},}\ }in\ \href
  {\doibase 10.1109/PAC.2007.4440844} {\emph {\bibinfo {booktitle} {2007 IEEE
  Particle Accelerator Conference (PAC)}}}\ (\bibinfo {year} {2007})\ pp.\
  \bibinfo {pages} {1625--1627}\BibitemShut {NoStop}%
\bibitem [{\citenamefont {B{\"o}hlen}\ \emph {et~al.}(2014)\citenamefont
  {B{\"o}hlen}, \citenamefont {Cerutti}, \citenamefont {Chin}, \citenamefont
  {Fass\`o}, \citenamefont {Ferrari}, \citenamefont {Ortega}, \citenamefont
  {Mairani}, \citenamefont {Sala}, \citenamefont {Smirnov},\ and\ \citenamefont
  {Vlachoudis}}]{FLUKA_Code}%
  \BibitemOpen
  \bibfield  {author} {\bibinfo {author} {\bibfnamefont {T.T.}\ \bibnamefont
  {B{\"o}hlen}}, \bibinfo {author} {\bibfnamefont {F.}~\bibnamefont {Cerutti}},
  \bibinfo {author} {\bibfnamefont {M.P.W.}\ \bibnamefont {Chin}}, \bibinfo
  {author} {\bibfnamefont {A.}~\bibnamefont {Fass\`o}}, \bibinfo {author}
  {\bibfnamefont {A.}~\bibnamefont {Ferrari}}, \bibinfo {author} {\bibfnamefont
  {P.G.}\ \bibnamefont {Ortega}}, \bibinfo {author} {\bibfnamefont
  {A.}~\bibnamefont {Mairani}}, \bibinfo {author} {\bibfnamefont {P.R.}\
  \bibnamefont {Sala}}, \bibinfo {author} {\bibfnamefont {G.}~\bibnamefont
  {Smirnov}}, \ and\ \bibinfo {author} {\bibfnamefont {V.}~\bibnamefont
  {Vlachoudis}},\ }\bibfield  {title} {\enquote {\bibinfo {title} {The {FLUKA
  Code}: Developments and challenges for high energy and medical
  applications},}\ }\href {\doibase https://doi.org/10.1016/j.nds.2014.07.049}
  {\bibfield  {journal} {\bibinfo  {journal} {Nuclear Data Sheets}\ }\textbf
  {\bibinfo {volume} {120}},\ \bibinfo {pages} {211 -- 214} (\bibinfo {year}
  {2014})}\BibitemShut {NoStop}%
\bibitem [{\citenamefont {Ahdida}\ \emph {et~al.}(2018)\citenamefont {Ahdida},
  \citenamefont {Calviani}, \citenamefont {Goddard}, \citenamefont
  {Jacobsson},\ and\ \citenamefont {Lamont}}]{Ahdida:2650896}%
  \BibitemOpen
  \bibfield  {author} {\bibinfo {author} {\bibfnamefont {C.}~\bibnamefont
  {Ahdida}}, \bibinfo {author} {\bibfnamefont {M.}~\bibnamefont {Calviani}},
  \bibinfo {author} {\bibfnamefont {B.}~\bibnamefont {Goddard}}, \bibinfo
  {author} {\bibfnamefont {R.}~\bibnamefont {Jacobsson}}, \ and\ \bibinfo
  {author} {\bibfnamefont {M.}~\bibnamefont {Lamont}},\ }\href
  {https://cds.cern.ch/record/2650896} {\emph {\bibinfo {title} {{SPS Beam Dump
  Facility Comprehensive Design Study}}}},\ \bibinfo {type} {Tech. Rep.}\
  (\bibinfo  {institution} {CERN},\ \bibinfo {address} {Geneva},\ \bibinfo
  {year} {2018})\BibitemShut {NoStop}%
\bibitem [{\citenamefont {Guarino}\ \emph {et~al.}(2001)\citenamefont
  {Guarino}, \citenamefont {Hauviller},\ and\ \citenamefont
  {Tavlet}}]{Radiation_instru}%
  \BibitemOpen
  \bibfield  {author} {\bibinfo {author} {\bibfnamefont {F.}~\bibnamefont
  {Guarino}}, \bibinfo {author} {\bibfnamefont {C.}~\bibnamefont {Hauviller}},
  \ and\ \bibinfo {author} {\bibfnamefont {M.}~\bibnamefont {Tavlet}},\ }\href
  {\doibase 10.5170/CERN-2001-006} {\emph {\bibinfo {title} {{Compilation of
  radiation damage test data. Adhesives}}}},\ CERN Yellow Reports: Monographs\
  (\bibinfo  {publisher} {CERN},\ \bibinfo {address} {Geneva},\ \bibinfo {year}
  {2001})\BibitemShut {NoStop}%
\bibitem [{\citenamefont {Liu}\ \emph {et~al.}(2016)\citenamefont {Liu},
  \citenamefont {Blokland}, \citenamefont {Bryan}, \citenamefont {Rakhman},
  \citenamefont {Riemer}, \citenamefont {Sangrey}, \citenamefont {Strum},
  \citenamefont {Wendel},\ and\ \citenamefont {Winder}}]{Straingauge_SNS}%
  \BibitemOpen
  \bibfield  {author} {\bibinfo {author} {\bibfnamefont {Y.}~\bibnamefont
  {Liu}}, \bibinfo {author} {\bibfnamefont {W.}~\bibnamefont {Blokland}},
  \bibinfo {author} {\bibfnamefont {J.}~\bibnamefont {Bryan}}, \bibinfo
  {author} {\bibfnamefont {A.}~\bibnamefont {Rakhman}}, \bibinfo {author}
  {\bibfnamefont {B.}~\bibnamefont {Riemer}}, \bibinfo {author} {\bibfnamefont
  {R.}~\bibnamefont {Sangrey}}, \bibinfo {author} {\bibfnamefont
  {R.}~\bibnamefont {Strum}}, \bibinfo {author} {\bibfnamefont
  {M.}~\bibnamefont {Wendel}}, \ and\ \bibinfo {author} {\bibfnamefont
  {D.}~\bibnamefont {Winder}},\ }\bibfield  {title} {\enquote {\bibinfo {title}
  {{Radiation-Resistant Fiber Optic Strain Sensors for SNS Target
  Instrumentation}},}\ }in\ \href {\doibase 10.18429/JACoW-IPAC2016-MOPMR055}
  {\emph {\bibinfo {booktitle} {{Proceedings, 7th International Particle
  Accelerator Conference (IPAC 2016): Busan, Korea, May 8-13, 2016}}}}\
  (\bibinfo {year} {2016})\ p.\ \bibinfo {pages} {MOPMR055}\BibitemShut
  {NoStop}%
\bibitem [{\citenamefont {\textit{et
  al.}}(2019{\natexlab{c}})}]{PhysRevAccelBeams.22.041001}%
  \BibitemOpen
  \bibfield  {author} {\bibinfo {author} {\bibfnamefont {N.~Simos}\
  \bibnamefont {\textit{et al.}}},\ }\bibfield  {title} {\enquote {\bibinfo
  {title} {{120 GeV neutrino physics graphite target damage assessment using
  electron microscopy and high-energy x-ray diffraction}},}\ }\href {\doibase
  10.1103/PhysRevAccelBeams.22.041001} {\bibfield  {journal} {\bibinfo
  {journal} {Phys. Rev. Accel. Beams}\ }\textbf {\bibinfo {volume} {22}},\
  \bibinfo {pages} {041001} (\bibinfo {year} {2019}{\natexlab{c}})}\BibitemShut
  {NoStop}%
\bibitem [{\citenamefont {Mcclintock}\ \emph {et~al.}(2014)\citenamefont
  {Mcclintock}, \citenamefont {Vevera}, \citenamefont {Riemer}, \citenamefont
  {Gallmeier}, \citenamefont {Hyres},\ and\ \citenamefont
  {Ferguson}}]{MCCLINTOCK2014130}%
  \BibitemOpen
  \bibfield  {author} {\bibinfo {author} {\bibfnamefont {D.}~\bibnamefont
  {Mcclintock}}, \bibinfo {author} {\bibfnamefont {B.}~\bibnamefont {Vevera}},
  \bibinfo {author} {\bibfnamefont {B.}~\bibnamefont {Riemer}}, \bibinfo
  {author} {\bibfnamefont {F.}~\bibnamefont {Gallmeier}}, \bibinfo {author}
  {\bibfnamefont {J.}~\bibnamefont {Hyres}}, \ and\ \bibinfo {author}
  {\bibfnamefont {P.}~\bibnamefont {Ferguson}},\ }\bibfield  {title} {\enquote
  {\bibinfo {title} {{Post-irradiation tensile properties of the first and
  second operational target modules at the Spallation Neutron Source}},}\
  }\href {\doibase 10.1016/j.jnucmat.2014.02.037} {\bibfield  {journal}
  {\bibinfo  {journal} {Journal of Nuclear Materials}\ }\textbf {\bibinfo
  {volume} {450}},\ \bibinfo {pages} {130--140} (\bibinfo {year}
  {2014})}\BibitemShut {NoStop}%
\end{thebibliography}%

\end{document}